\documentclass[lettersize,journal]{IEEEtran}
\usepackage{amsmath,amsfonts}
\usepackage{algorithmic}
\usepackage{algorithm}
\usepackage{array}
\usepackage[caption=false,font=normalsize,labelfont=sf,textfont=sf]{subfig}
\usepackage{textcomp}
\usepackage{stfloats}
\usepackage{url}
\usepackage{verbatim}
\usepackage{graphicx}
\usepackage{cite}
\begin{document}

\title{Optimized Design of the Generalized Bilinear Transformation for Discretizing Analog Systems}

\author{Shen Chen,~\IEEEmembership{Student Member,~IEEE,} Yanlong Li, Jiamin Cui, Wei Yao, 
Jisong Wang, Yixin Tian, Chaohou Liu, Yang Yang, Jiaxi Ying, Zeng Liu,~\IEEEmembership{Senior Member,~IEEE,} Jinjun Liu,~\IEEEmembership{Fellow,~IEEE}
\thanks{ Manuscript received November 5, 2025.
This work was supported by the National Key Research and Development Program of China under Grant 2023YFB2604600.
(Corresponding author: Jinjun Liu.)

Shen Chen is with the School of Electrical Engineering, Xi'an Jiaotong University, Xi'an 710049, China,
and also with SolaX Power Network Technology (Zhejiang) Co., Ltd., Hangzhou 310013, China 
(e-mail: chenshen@stu.xjtu.edu.cn).

Zeng Liu, and Jinjun Liu are with the State Key Laboratory of Electrical Insulation and
Power Equipment, School of Electrical Engineering, Xi'an Jiaotong University, Xi'an 710049, China 
(e-mail: zengliu@mail.xjtu.edu.cn; jjliu@mail.xjtu.edu.cn).

Yanlong Li, Jisong Wang, Yixin Tian, Chaohou Liu, Yang Yang, and Jiaxi Ying are with
SolaX Power Network Technology (Zhejiang) Co., Ltd., Hangzhou 310013, China 
(e-mail: liyanlong@solaxpower.com; wangjisong@solaxpower.com; 
tianyixin@solaxpower.com; liuchaohou@solaxpower.com; yangyang@solaxpower.com; yingjiaxi@solaxpower.com).

Jiamin Cui and Wei Yao are with the Key Laboratory of Technology in Rural Water Management of Zhejiang Province, 
College of Electric Engineering, Zhejiang University of Water Resources and Electric Power, Hangzhou 310018, China
(e-mail: yaowei@zjweu.edu.cn; cuijm@zjweu.edu.cn).
}
}

\markboth{IEEEtran \LaTeX\ Template v1.8b,~Vol.~x, No.~x, October~2025}%
{Shell \MakeLowercase{\textit{et al.}}: A Sample Article Using IEEEtran.cls for IEEE Journals}


\maketitle

\begin{abstract}
A common approach to digital system design involves transforming a continuous-time (s-domain) transfer function 
into the discrete-time (z-domain) using methods such as Euler or Tustin.
These transformations are shown to be specific cases of the Generalized Bilinear Transformation (GBT), characterized by a design parameter, $\alpha$,
whose physical interpretation and optimal selection remain inadequately explored.
In this paper, we propose an alternative derivation of the GBT derived by employing a new hexagonal shape 
to approximate the enclosed area of the error function, and we define the parameter $\alpha$ as a shape factor.
We reveal, for the first time, the physical meaning of $\alpha$ as the backward rectangular ratio of the proposed hexagonal shape. 
Through domain mapping, the stable range of is rigorously established to be [0.5, 1].
Depending on the operating frequency and the chosen $\alpha$, 
we observe two distinct distortion modes, i.e., the magnitude and phase distortion.
We further develop an optimal design method for $\alpha$ by minimizing a normalized magnitude or phase error objective function.
The effectiveness of the proposed method is validated through the design and testing of a low-pass filter (LPF), 
demonstrating strong agreement between theoretical predictions and experimental results.
\end{abstract}

\begin{IEEEkeywords}
Discretization, generalized bilinear transformation, numerical integration, hexagonal approximation, shape factor, distortion, optimal design.
\end{IEEEkeywords}

\section{Introduction}
\IEEEPARstart{D}{igital} control technology has revolutionized modern industrial systems, 
and has become the cornerstone of automation in diverse fields such as manufacturing, robotics, and engineering \cite{Fadali13-DCE}.
By optimizing the real-time signal chain \cite{TIAN-SPRACW5A}, 
the digitally-controlled systems achieve high accuracy, flexibility, and consistency in processes that were previously dominated by analog systems.
An essential step in implementing digital systems is the discretization.
Generally, there are two broad approaches to implementing discretization. 
The first approach is the direct discrete design \cite{Direct-01} --\cite{Direct-05}, which is applied directly to a discrete control plant.
The second approach is the indirect discrete design \cite{Euler-01} --\cite{GBT-05}, which involves designing the analog systems in the continuous-time domain (s-domain)
and subsequently transforming them into the discrete-time domain (z-domain) using the s-to-z transformation.
However, all discretization methods introduce undesired errors including magnitude and phase distortion.
Furthermore, some methods may even cause the discrete system to become unstable.
Therefore, the selection of an appropriate discretization methods is crucial during the digital implementation process.

\begin{figure}[h] 
  \centering
  \includegraphics[width=1.0\linewidth]{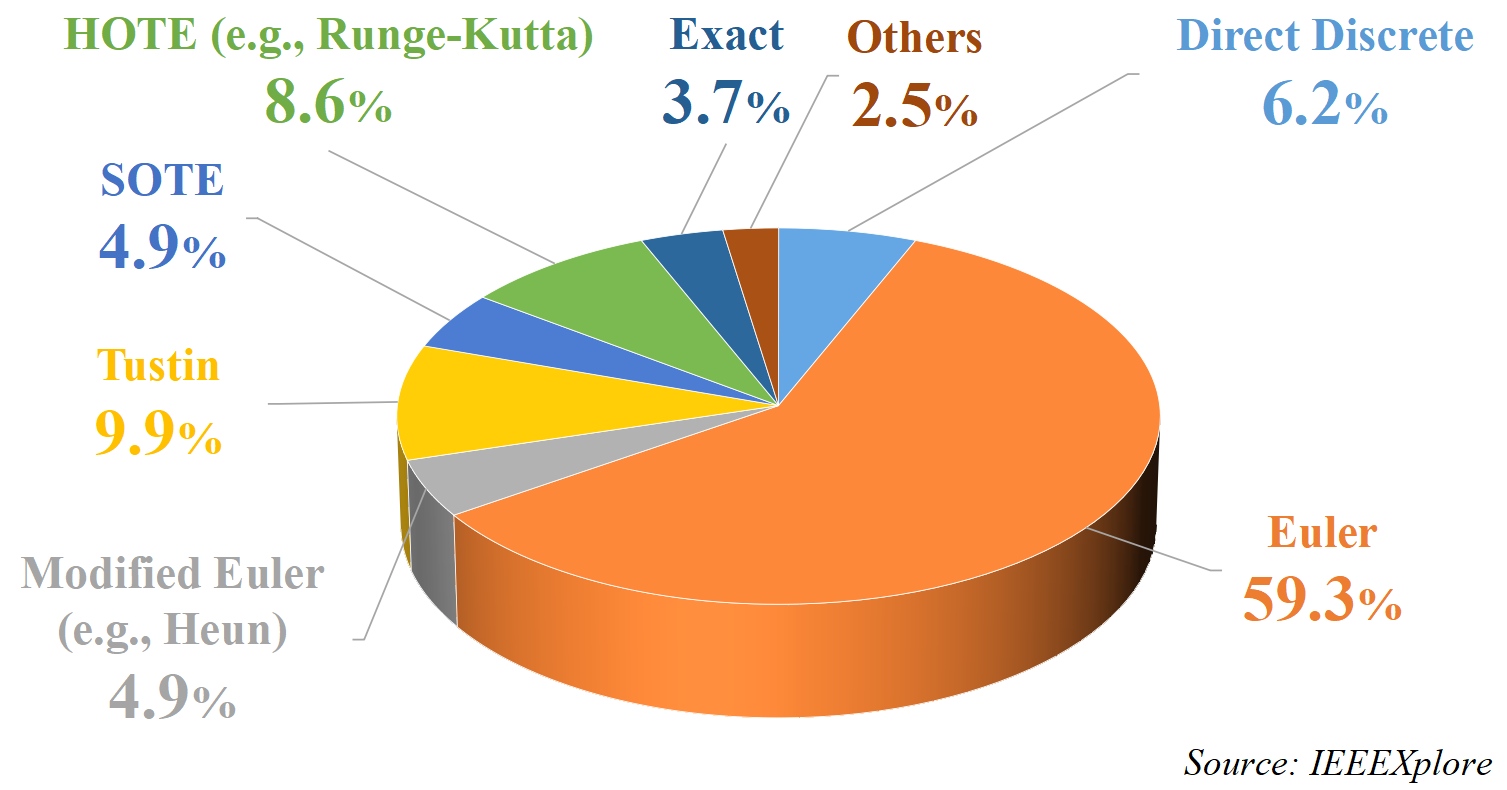}
  \caption{Distribution graph of discretization methods based on 81 papers(\cite{Direct-01} -- \cite{Others-02}) published with the keyword "discretization" in "\textit{IEEE Trans. Power Electron.}", "\textit{IEEE Trans. Ind. Inform.}", and "\textit{IEEE Trans. Ind. Electron.}" since 2023. 
  SOTE: Second Order Taylor Expansion; HOTE: Higher Order Taylor Expansion.}
  \label{fig_method_distribution}
\end{figure}

Although the direct approach has become popular for modern control theory, 
the traditional indirect approach remains the dominant method in industrial applications.
As shown in Fig.~\ref{fig_method_distribution}, 
about 93.8\% of publications in this review (see the Appendix for further details) employ the indirect approach (excluding the "Direct Discrete").
When considering the trade-off between computational effort and accuracy, 
the Euler method (including the backward and forward variants) \cite{Euler-01} --\cite{Euler-48} is the most widely adopted form of the indirect approach. 
However, the discretization error becomes unacceptably large near the Nyquist frequency.
In such cases, the Tustin method (also referred to as the "bilinear" method) \cite{Tustin-01} --\cite{Tustin-08} demonstrates superior performance 
due to its phase response matching that of the exact discretization \cite{GBT-05}.
The formulations of these two methods are provided in equations \eqref{eq1} and \eqref{eq2}, respectively:
\begin{equation} s = \frac{2}{T}\frac{z-1}{z+1}\label{eq1}\end{equation}
\begin{equation} s = \frac{1}{T}\frac{z-1}{z}\label{eq2}\end{equation}
where $T$ is the sampling period.
However, the Tustin method introduces a frequency-warping phenomenon near the Nyquist frequency.
To mitigate this issue, the frequency pre-warping method is applied:
\begin{equation} \omega_{pwp}=\frac{2}{T}tan(\frac{\omega_{ori}\cdot T}{2})\label{eq_prewarp}\end{equation}
where $\omega_{ori}$ and $\omega_{pwp}$ represent the original and pre-warped frequencies of the analog system, respectively.
In applications demanding high-precision control, 
such as Phase-Locked Loop (PLL) dynamics in weak grids \cite{SOTE-01}, surgical robot control \cite{HOTE-02}, 
and PMSM control under low frequency ratios \cite{HOTE-RK-01}, 
the truncation errors inherent in both the Euler and Tustin methods (derived from first-order approximations) can degrade control performance.
For such scenarios, higher-order discretization methods are recommended, 
including the second-order Taylor expansion method \cite{SOTE-01} --\cite{SOTE-03} and the fourth-order Runge-Kutta method \cite{HOTE-RK-01} --\cite{HOTE-RK-03}. 
For instance, \cite{SOTE-01} employs a second-order sliding mode differentiator (SOSMD) 
to enhance PLL dynamics in grid-connected inverters under weak grid conditions. 
Compared to differentiators based on the Euler or Tustin methods, 
the SOSMD offers improved finite-time convergence and stronger disturbance robustness.
However, the complexity and computation cost increase.

In practice, the Euler and the Tustin methods 
are two of the most commonly used indirect methods favored for their simplicity and practicality.
As illustrated in Fig.~\ref{fig_method_distribution}, the Euler method (including its modified variants) accounts for approximately 64.2\% of applications, 
while the Tustin method represents about 9.9\%.
Moreover, these two methods can be unified under the framework of the Generalized Bilinear Transformation (GBT).
The concept of the GBT was first introduced by Sekera in 2005 as the $\alpha$-approximation \cite{GBT-01-Sekara}.
This method is derived from the first-order approximations of both the numerator and the denominator, formulated as follows:
\begin{equation} s = \frac{1}{T}\frac{z-1}{\alpha z+(1-\alpha)}\label{eq_gbt}\end{equation}
where $\alpha$ is a design parameter within the range [0,1].
However, the physical meaning of the parameter $\alpha$ remains unclear, 
and no theoretical analysis of discretization error for different $\alpha$ was provided.
In 2008, the Al-Alaoui integrator {\cite{GBT-02-Alaoui}} was proposed:
\begin{equation} s = \frac{2}{T}\frac{z-1}{(1+a)z+(1-a)}\label{eq3}\end{equation}
where $a$ is a design parameter within the range [0,1].
This formulation interpolates the trapezoidal and the rectangular integration rules.
Notably, \cite{GBT-03} demonstrates that the GBT and the Al-Alaoui operator are mathematically identical.
Despite this, both approaches lack a physical explanation for the design parameters and their impact on the discretization error.
\cite{GBT-04} extends the GBT by expanding the range of the design parameter from [0,1] into $(-\infty, \infty)$,
offering a broader class of digital approximations for analog systems.
However, this work overlooks stability issue: when $\alpha$ exceeds certain bounds, the transformed system may become unstable in the z-domain.
\cite{GBT-05} presents an accurate discretization method,
\begin{equation} s = \frac{1+\alpha_p}{T}\frac{z-1}{z+\alpha_p}\label{eq5}\end{equation}
where $\alpha_p$ is a design parameter within the range [0,1].
While this method is equivalent to the GBT, it similarly fails to clarify the physical meaning of $\alpha_p$ or provide guidance for its optimal design.

To the best of author's knowledge, 
existing literature has not elucidated the physical meaning and optimal design method for the parameter $\alpha$ in the GBT. 
To address this gap, this study aims to clarify the physical meaning of $\alpha$, analyze its impact on discretization error,
and develop an optimal design method for $\alpha$ selection.
We demonstrate that the GBT serves as a unified framework for numerical integration
and provides a tunable degree of freedom to regulate discretization error.
Notably, when $\alpha$ is set to 0.5 or 1, the GBT turns into the Tustin or the Euler method, respectively. 
The paper is organized as follows:
First, we present a novel mathematical derivation of the GBT and its relationship with existing variations.
Next, we conduct the stability analysis based on domain mapping.
Additionally, we explore the discretization error in terms of magnitude and phase distortion
by analyzing how the Bode plot changes under different shape factors.
Then, we propose an optimal design method for the shape factor, $\alpha$.
Finally, we use the GBT to discretize a low-pass filter (LPF).
The proposed method is validated by comparing the theoretical predictions with the experimental results.

\section{Novel Derivation of the GBT}
\subsection{Conventional Derivation of the GBT}
In the process of discretization of analog systems, the exact mapping from the s-domain to the z-domain is denoted as $z = e^{sT}$.
This transformation maps the left half of the s-plane to the interior of the unit circle in the z-plane.
Starting from this basic transformation, the equivalent relation is defined as follows: 
\begin{equation} z = e^{sT} = e^{s[(1-\alpha)T+\alpha T]} = \frac{e^{s(1-\alpha)T}}{e^{-s\alpha T}}, \alpha \in [0,1]\label{eq_alpha_appr}\end{equation}

Using the Taylor expansion for both the numerator and the denominator on the right side of expression \eqref{eq_alpha_appr} and neglecting all terms of second order and higher,
we obtain:
\begin{equation} z = \frac{\sum_{n=0}^{\infty}\frac{[s(1-\alpha)T]^n}{n!}}{\sum_{m=0}^{\infty}(-1)^k \frac{(s\alpha T)^m}{m!}} \approx \frac{1+s(1-\alpha)T}{1-s\alpha T}\label{eq_alpha_appr2}\end{equation}

Solving equation \eqref{eq_alpha_appr2} for the variable $s$ yields the first-order approximation, which is defined as GBT.
\begin{equation} s = \frac{1}{T}\frac{z-1}{1+\alpha(z-1)}\label{eq_alpha_appr3}\end{equation}
where $\alpha$ is a design parameter with the range [0, 1].

\subsection{Novel Derivation Based on Numerical Integration}
\begin{figure}[h]  
  \centering
  \includegraphics[width=1.0\linewidth]{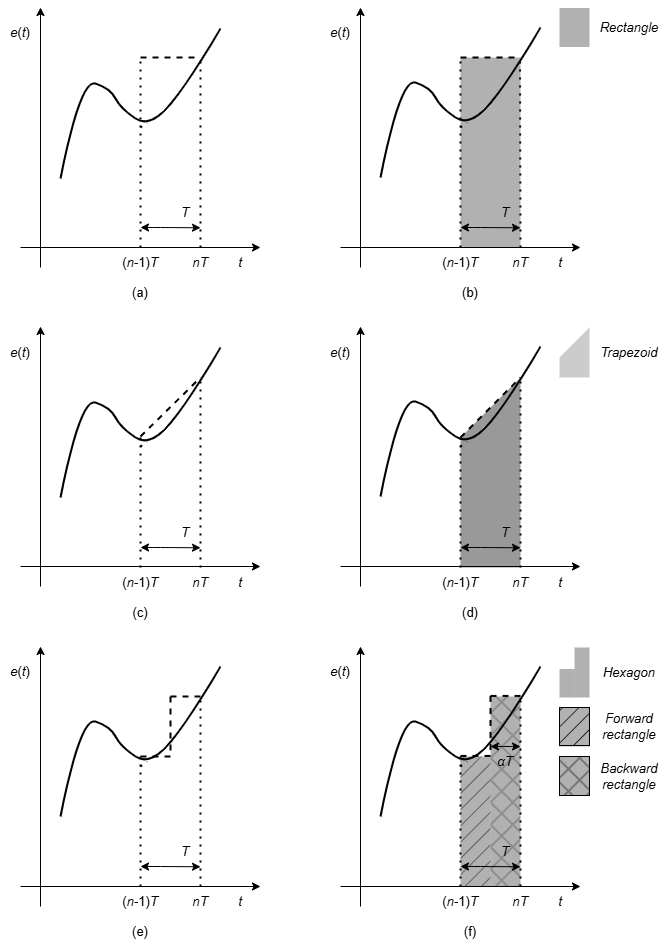}
  \caption{Mathematical interpretation. (a) Euler method. (b) Rectangular approximation. (c) Tustin method. (d) Trapezoidal approximation. (e) GBT method. (f) Hexagonal approximation.}
  \label{fig_approximation}
\end{figure}

Starting from the relationship between the error function, $e(t)$, and the original function, $u(t)$, as illustrated below:
\begin{equation} e(t) = \frac{d u(t)}{d t} \label{eq_diff}\end{equation}

In the continuous domain, $u(t)$ is expressed as follows: 
\begin{equation} u(t) = \int e(t) \,dt  \label{eq_u_e_cont}\end{equation}

In the discrete domain, $u(n)$ is expressed as follows:
\begin{equation} u(n) = \int_{(n-1)T}^{nT} e(t) \,dt + u(n-1) \label{eq_u_e_disc}\end{equation}

Fig.~\ref{fig_approximation} compares the mathematical interpretation of the GBT and other methods.
Fig.~\ref{fig_approximation}(a) illustrates the Euler method geometrically.
The area enclosed by the three dotted lines and the horizontal axis is rectangular.
This rectangular approximation equals the numerical integration of the solid area as illustrated in Fig.~\ref{fig_approximation}(b).
In this case, the error function $e(t)$ is expressed as follows:
\begin{equation} e(t) = e(n) \label{eq_rect_error}\end{equation}
In the discrete domain, $u(n)$ is expressed as follows:
\begin{equation} u(n) = \int_{(n-1)T}^{nT} e(t) \,dt + u(n-1) = e(n)\cdot T + u(n-1) \label{eq_u_rect_appr}\end{equation}

Fig.~\ref{fig_approximation}(c) illustrates the Tustin method geometrically.
The area enclosed by the three dotted lines and the horizontal axis forms a trapezoid.
This trapezoidal approximation equals the numerical integration of the solid area as illustrated in Fig.~\ref{fig_approximation}(d).
In this case, the error function $e(t)$ is expressed as follows: 
\begin{equation} 
  \begin{split}
      e(t) &= \frac{e(n)-e(n-1)}{T} \cdot [t-(n-1)T] + e(n-1) \\  
            & ,(n-1)T < t < n\cdot T      
  \end{split}
  \label{eq_trap_error}
\end{equation}
In the discrete domain, $u(n)$ is expressed as follows:
\begin{equation} u(n) = \frac{e(n)+e(n-1)}{2} \cdot T + u(n-1) \label{eq_u_trap_appr}\end{equation}

Fig.~\ref{fig_approximation}(e) illustrates the GBT geometrically.
The area enclosed by the five dotted lines and the horizontal axis forms a hexagon, 
which is the reason this approach is referred to as the hexagonal approximation. 
This hexagonal approximation is equivalent to the numerical integration of the solid area shown in Fig.~\ref{fig_approximation}(f).
In this case, error function $e(t)$ is expressed as follows:
\begin{equation} 
  e(t) = 
  \begin{cases}
      e(n-1), & t \in [(n-1)T, (n-\alpha)T] \\
      e(n), & t \in ((n-\alpha)T, n\cdot T]
  \end{cases}
  \label{eq_hexa_error}
\end{equation}
In the discrete domain, $u(n)$ is expressed as follows:
\begin{equation} u(n) = (1-\alpha)\cdot e(n-1)T + \alpha \cdot e(n)T + u(n-1) \label{eq_u_hexa_appr}\end{equation}
Therefore,
\begin{equation} (1 - z^{-1})\cdot U(z) = [(1-\alpha)\cdot z^{-1}\cdot T + \alpha\cdot T] \cdot E(z)\label{eq_Uz_Ez}\end{equation}
\begin{equation} s = \frac{E(z)}{U(z)} = \frac{1}{T}\frac{1-z^{-1}}{\alpha + (1-\alpha) \cdot z^{-1}}\label{eq_s2z_hexa}\end{equation}

Moreover, the hexagonal area in Fig.~\ref{fig_approximation}(f) comprises two rectangular parts.
The left part is a forward rectangular area and the right part is a backward rectangular area.
The physical meaning of the parameter $\alpha$ is the percentage of the backward rectangular area as defined in equation \eqref{eq_alpha_def}.

\begin{equation} \alpha = \frac{S_{bw\_rec}}{S_{bw\_rec} + S_{fw\_rec}}\label{eq_alpha_def}\end{equation}
where $S_{bw\_rec}$ and $S_{fw\_rec}$ are the backward rectangular area and the forward rectangular area, respectively.

\subsection{Relations with the Existing Methods}
The relationship between the GBT and some existing methods is shown in Table~\ref{tab:relation}. 
The Euler and the Tustin methods are two specific forms of GBT, with $\alpha$ equals 1 and 0.5, respectively.
The Al-Alaoui operator \cite{GBT-02-Alaoui} is equivalent to the GBT. 
The difference is the design parameters, and their relationship is given by $\alpha=\frac{1+a}{2}$.
In this case, $\alpha$ is limited to [0.5, 1] because the range of the parameter $a$ is [0, 1].
This result is also consistent with the nature of the Al-Alaoui operator, 
which interpolates the trapezoidal (Tustin) and the rectangular (Euler) integration rules.
\cite{GBT-04} is equivalent to the GBT because the design parameters are similar.
The only difference is in the range of the design parameter: \cite{GBT-04} extends the range of $\alpha$ from [0,1] to $(-\infty, \infty)$.
The method in \cite{GBT-05} is also equivalent to the GBT. 
The only difference lies in the parameters, and their relationship is expressed as $\alpha=\frac{1}{1+\alpha_p}$.

\begin{table}[!t] 
\caption{Relations with the Existing Methods\label{tab:relation}}
\centering
\begin{tabular}{|c|c|c|c|}
\hline
\textbf{Method} & \textbf{Transformation} & \textbf{Range} & \textbf{Relationship}  \\
\hline
GBT in \cite{GBT-01-Sekara}    & $s = \frac{1}{T}\frac{z-1}{\alpha z+(1-\alpha)}$      & [0,1]               & \textbf{reference} \\
Euler method                   & $s = \frac{1}{T}\frac{z-1}{z}$                        & /                   & $\alpha$=1 \\
Tustin method                  & $s = \frac{2}{T}\frac{z-1}{z+1}$                      & /                   & $\alpha$=0.5\\
Al-Alaoui \cite{GBT-02-Alaoui} & $s = \frac{2}{T}\frac{z-1}{(1+a)z+(1-a)}$             & [0,1]               & $\alpha=\frac{1+a}{2}$\\
GBT in \cite{GBT-04}           & $s = \frac{1}{T}\frac{z-1}{\alpha_g z+(1-\alpha_g)}$  & $(-\infty, \infty)$ & $\alpha=\alpha_g$\\
Method in \cite{GBT-05}        & $s = \frac{1+\alpha_p}{T}\frac{z-1}{z+\alpha_p}$      & [0,1]               & $\alpha=\frac{1}{1+\alpha_p}$\\
\hline
\end{tabular}
\end{table}

\section{Comprehensive Analysis}\label{sec3}
\subsection{Stability Analysis}\label{sub1sec3}
\begin{figure}[h] 
  \centering
  \includegraphics[width=1.0\linewidth]{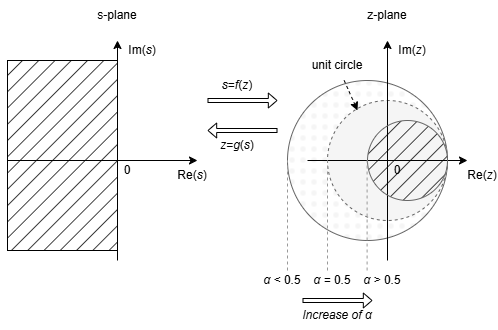}
  \caption{Mapping of s-plane to z-plane}
  \label{fig_s2z_mapping}
\end{figure}

Let $s=\sigma_s + j\omega_s$ and $z=\gamma_z + j\cdot \zeta_z $, we have the following expression by substituting these into equation \eqref{eq_gbt}:
\begin{equation}   
  \begin{split}
      s &= \sigma_s + j\omega_s = \frac{1}{T}\frac{[\gamma_z + j\cdot \zeta_z]-1}{\alpha (\gamma_z + j\cdot \zeta_z)+(1-\alpha)} \\  
         &=\frac{1}{T}\frac{[(\gamma_z-1)+j\cdot \zeta_z][(\alpha \gamma_z+1-\alpha)-j\alpha\cdot \zeta_z]}{[\alpha \gamma_z+1-\alpha]^2+[\alpha\cdot \zeta_z]^2}  \\    
         &=\frac{1}{T}\frac{[\alpha(\gamma_z-1)^2+\gamma_z-1+\alpha\cdot \zeta_z^2]+j\alpha\cdot \zeta_z}{[\alpha \gamma_z+1-\alpha]^2+[\alpha\cdot \zeta_z]^2}    \\    
  \end{split}
  \label{eq_stab1}
\end{equation}
Therefore, $\sigma_s$ and $\omega_s$ can be derived as follows:
\begin{equation} \sigma_s = \frac{1}{T}\frac{\alpha(\gamma_z-1)^2+\gamma_z-1+\alpha\cdot \zeta_z^2}{[\alpha \gamma_z+1-\alpha]^2+[\alpha\cdot \zeta_z]^2} \label{eq_sigma_s}\end{equation}
\begin{equation} \omega_s = \frac{1}{T}\frac{\zeta_z}{[\alpha \gamma_z+1-\alpha]^2+[\alpha\cdot \zeta_z]^2} \label{eq_omega_s}\end{equation}
As previously mentioned, the transformation is stable unless the left half of the s-plane is mapped into the unit circle of the z-plane, which implies that $\sigma_s \leq  0$.
Substituting this into equation \eqref{eq_sigma_s} yields,
\begin{equation} \alpha(\gamma_z-1)^2+\gamma_z-1+\alpha\cdot \zeta_z^2 \leq 0\label{eq_stab2}\end{equation}
Therefore, we have:
\begin{equation} [\gamma_z - (1-\frac{1}{2\alpha})]^2 + \zeta_z^2 \leq (\frac{1}{2\alpha})^2\label{eq_stab_cond}\end{equation}
This equation has two crossing points on the real axis, labeled as $\gamma_{z1}$ and $\gamma_{z2}$, respectively,
\begin{equation} \gamma_{z1} = 1, \gamma_{z2} = 1-\frac{1}{\alpha}\label{eq_stab4}\end{equation}
Since $z$ should be within the unit circle of the z-plane, therefore, we have:
\begin{equation} \gamma_{z2} = 1-\frac{1}{\alpha} \geq -1\label{eq_stab5}\end{equation} 
Therefore, the restriction for a stable transformation can be expressed as $0.5 \leq \alpha \leq 1.0$.

Fig.~\ref{fig_s2z_mapping} shows clearly that the mapping is stable when $\alpha \geq 0.5$. 
In conclusion, the shape factor $\alpha$ should be within the range [0.5, 1] in order to achieve a stable transformation. 
Specifically, when $\alpha$ is set to 0.5 or 1, the GBT turns into the Tustin or the Euler method, respectively. 

\subsection{Distortion Analysis}\label{sub2sec3}
\begin{figure*}[h]  
  \centering
  \includegraphics[width=1.0\linewidth]{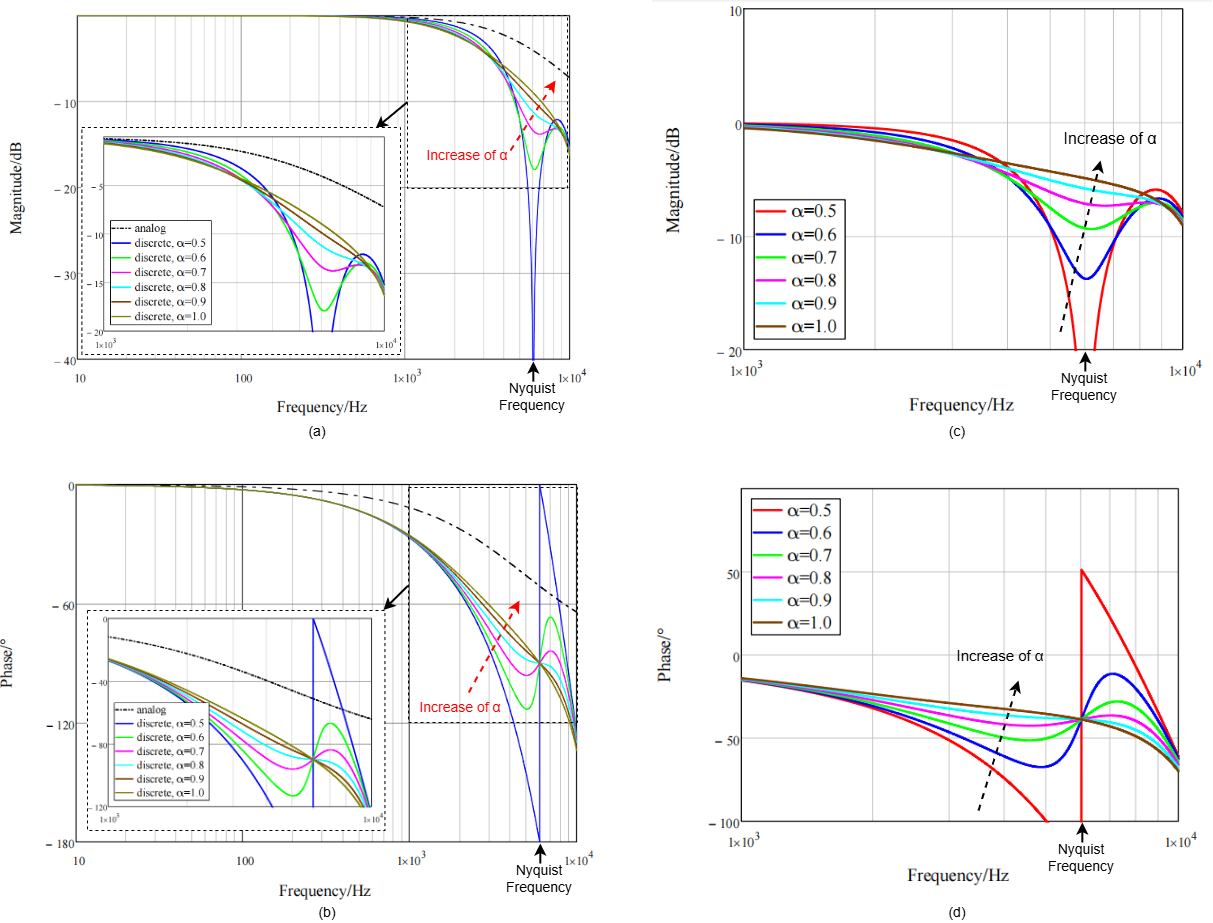}
  \caption{Magnitude and phase distortion of discrete LPF with different $\alpha$}
  \label{fig_distortion}
\end{figure*}

For an analog system with $m$ zeros and $n$ poles, the generalized form of the transfer function in the s-domain is given by:
\begin{equation} G_{anlg}(s) = K \frac{\sum_{i=1}^{m}(s+\mathbf{Z_i})}{\sum_{k=1}^{n}(s+\mathbf{P_k})}\label{eq_analog_plant}\end{equation} 
where $K$ is the system gain, $\mathbf{Z_i}$ is the $i$-th zero, $\mathbf{P_k}$ is the $k$-th pole.

The frequency response in the s-domain is denoted by:
\begin{equation} G_{anlg}(w_s) = K \frac{\sum_{i=1}^{m}(j\cdot w_s+\mathbf{Z_i})}{\sum_{k=1}^{n}(j\cdot w_s+\mathbf{P_k})}\label{eq_plant_s_freq_resp}\end{equation} 
where $\omega_s$ is the angular frequency in the s-domain, $j$ is the imaginary unit.

Substituting equation \eqref{eq_gbt} into equation \eqref{eq_analog_plant} yields the transformed discrete system:
\begin{equation} G_{disc}(z,\alpha) = K \frac{\sum_{i=1}^{m}(\frac{1}{T}\frac{z-1}{\alpha z+(1-\alpha)}+\mathbf{Z_i})}{\sum_{k=1}^{n}(\frac{1}{T}\frac{z-1}{\alpha z+(1-\alpha)}+\mathbf{P_k})}\label{eq_discrete_plant}\end{equation} 

Considering the effect of zero-order-hold (ZOH) sampling, the frequency response in the z-domain is given by,
\begin{equation} 
    \begin{split}
    &G_{disc}(w_z,\alpha) = \\
    &K \frac{\sum_{i=1}^{m}(\frac{1}{T}\frac{e^{j w_z T}-1}{\alpha e^{j w_z T}+(1-\alpha)}+\mathbf{Z_i})}{\sum_{k=1}^{n}(\frac{1}{T}\frac{e^{j w_z T}-1}{\alpha e^{j w_z T}+(1-\alpha)}+\mathbf{P_k})}\frac{\sin(0.5 w_z T)}{j0.5 w_z T}e^{-j0.5 w_z T} = \\
    &K \frac{\sum_{i=1}^{m}(\frac{1}{T}\frac{(1-2\alpha)(cos(w_z T)-1)+j sin(w_z T)}{(2\alpha-2\alpha^2)cos(w_z T) + (2\alpha^2-2\alpha+1)}+\mathbf{Z_i})}{\sum_{k=1}^{n}(\frac{1}{T}\frac{(1-2\alpha)(cos(w_z T)-1)+j sin(w_z T)}{(2\alpha-2\alpha^2)cos(w_z T) + (2\alpha^2-2\alpha+1)}+\mathbf{P_k})} \cdot \\
                &\frac{\sin(0.5 w_z T)}{0.5 w_z T} \cdot e^{-j0.5 w_z T}\\
    \end{split}
    \label{eq_plant_z_freq_resp}
\end{equation} 
where $\omega_z$ is the angular frequency in the z-domain, 
$\frac{\sin(0.5 w_z T)}{0.5 w_z T}$ and $e^{-j0.5 w_z T}$ are auxiliary magnitude decay and phase delay induced by ZOH sampling, respectively. 

To study the influence of the shape factor $\alpha$ on the Bode plot, an LPF is introduced,
\begin{equation} G_{LPF}(s) = \frac{w_c}{s+w_c}\label{eq_lpf}\end{equation} 
where $w_c$ is the crossing angular frequency. 
The frequency responses of the LPF in the s-domain and z-domain are expressed as follows:
\begin{equation} G_{LPF\_anlg}(f) = \frac{w_c}{j2\pi f+w_c}\label{eq_lpf_analog}\end{equation} 
\begin{equation} 
  \begin{split}
  &G_{LPF\_disc}(f, \alpha) = \\
  &\frac{w_c}{\frac{1}{T}\frac{(1-2\alpha)(cos(2\pi f\cdot T)-1)+j sin(2\pi f\cdot T)}{(2\alpha-2\alpha^2)cos(2\pi f\cdot T) + (2\alpha^2-2\alpha+1)}+w_c}\frac{\sin(\pi f\cdot T)}{\pi f\cdot T} e^{-j\pi f\cdot T}\\
  \end{split}
  \label{eq_lpf_discrete}
\end{equation} 

The Bode plots of the discrete LPF with different $\alpha$ are illustrated in Fig.~\ref{fig_distortion}.
In this case, the sampling frequency is 12 kHz; therefore, the Nyquist frequency is 6 kHz. The LPF's crossing frequency is 4.823 kHz.
Since the GBT is a first-order approximation, it inevitably introduces distortion including magnitude and phase errors.
In this study, we observed two distinct distortion modes: i.e. magnitude distortion and phase distortion.
These are illustrated in Fig.~\ref{fig_distortion}(a) and Fig.~\ref{fig_distortion}(b), respectively.
Fig.~\ref{fig_distortion}(c) and Fig.~\ref{fig_distortion}(d) show how the magnitude and phase errors change 
when using the analog response as a reference and changing $\alpha$ from 0.5 to 1.0 in increments of 0.1.

\begin{figure}[h]  
  \centering
  \includegraphics[width=0.8\linewidth]{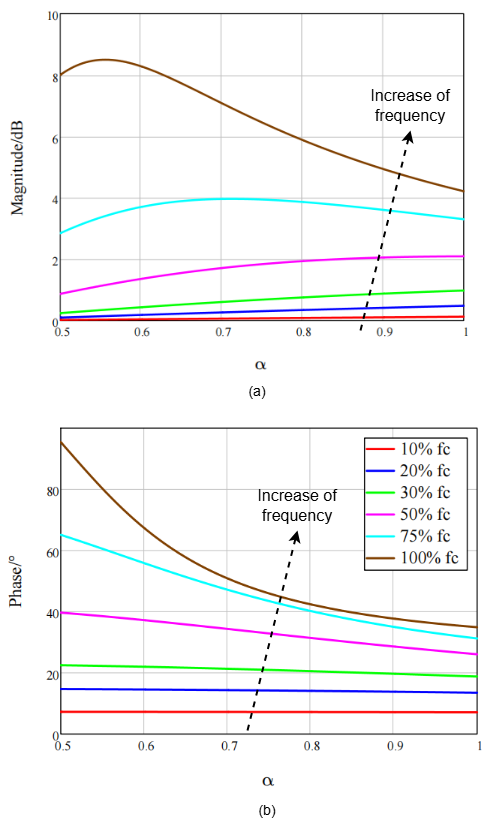}
  \caption{Magnitude and phase error vs. $\alpha$}
  \label{fig_err_vs_alpha}
\end{figure}

As shown in Fig.~\ref{fig_err_vs_alpha}, the magnitude and phase error vary with different frequencies and $\alpha$.
However, we perceive that there is an optimal shape factor, which achieves a minimum magnitude or phase error for a given frequency.

\section{Optimal Design Method and Application}\label{sec4}
In this section, we first propose an optimal design method for the shape factor $\alpha$ consisting of five steps.
Next, we use this method to implement discrete LPF.
\subsection{Optimal Design Method}\label{sub1sec4}
The proposed optimal design method for the shape factor $\alpha$ consists of five steps: 
application scenario, initialization, constraints, the objective function, and computation.

\textbf{Step 1: Application Scenario.}
Generally, there are three types of application scenarios based on the magnitude or phase optimization with frequency selection requirement.
\begin{enumerate}
  \item Type A: single frequency point. For this type, minimizing the normalized global error (defined as $Q_A(q, \alpha)$) 
  consisting of the magnitude and phase error is required in single frequency point (defined as $f_exp$).
  \item Type B: multiple frequency points. For this type, minimizing the normalized global error (defined as $Q_B(q, \alpha)$) is required, 
  which is the weighted cumulative error of multiple frequency points (defined as $\mathbf{f_{exp}}[1], \mathbf{f_{exp}}[2],..., \mathbf{f_{exp}}[N]$).
  \item Type C: frequency interval. For this type, 
  minimizing the normalized global error (defined as $Q_C(q, \alpha)$) is required, 
  which is the integral errors of an expected frequency interval (defined as $[f_{start}, f_{end}]$).
\end{enumerate}

\textbf{Step 2: Initialization.}
For model discretization, the Euler method, a special case of the GBT, is widely adopted due to its simplicity. 
This can be seen in applications such as permanent magnet synchronous motor (PMSM) discretization \cite{SOTE-01} - \cite{SOTE-03}.
For controller discretization, the Tustin method, another special case of the GBT, is often preferred in engineering practice. 
Owing to its excellent stability and relatively straightforward implementation, 
the Tustin method is considered as the preferred method for the general-purpose controllers \cite{Tustin-06}, 
such as the PI or proportional-resonant (PR) controllers.
Therefore, it is intuitively reasonable to set $\alpha$ to 0.5 or 1.0 because the Tustin and Euler methods are so commonly used.
However, it is recommended to use the random initialization method in order to avoid local-optimal trap.
Thus, the initial value of the shape factor $\alpha$ is defined by:
\begin{equation} \alpha_{init} = random(0.5, 1)\label{eq_alpha_init}\end{equation} 

\textbf{Step 3: Constraints.}
The stability of the discrete system is the most important criteria to judge the performance of optimized parameter,
therefore, the shape factor can't exceed the range of stability requirement during the process of computation.
For this reason, based on the conclusion of Section \ref{sub1sec3}, the constraints are given by:
\begin{equation} 0.5 \leq \alpha \leq 1 \label{eq_alpha_constraint}\end{equation} 

\textbf{Step 4: Objective Function. }
The magnitude error between the analog system and the discrete system is given by,
\begin{equation} L_{err}(f, \alpha) = 20\log(|\frac{G_{LPF\_disc}(f, \alpha)}{G_{LPF\_anlg}(f)}|)\label{eq_mag_err}\end{equation} 

The phase error between the analog system and the discrete system is given by,
\begin{equation} \phi_{err}(f, \alpha)=\angle(G_{LPF\_disc}(f, \alpha))-\angle(G_{LPF\_anlg}(f))\label{eq_phase_err}\end{equation} 

Based on the requirements of minimizing the magnitude or the phase error, the local error is denoted by,
\begin{equation} Q_{L}(f, \alpha) = \min(|L_{err}(f, \alpha)|)\label{eq_obj_func_minMag}\end{equation} 
or,
\begin{equation} Q_{\phi}(f, \alpha) = \min(|\phi_{err}(f, \alpha)|)\label{eq_obj_func_minPhase}\end{equation} 

The magnitude error and phase error are not in the same dimension, and cannot be compared directly.
Therefore, a normalization process is required. We use the MaxAbs scale method because it is a simple and practical feature-scaling method 
that is proved to be capable of improving the accuracy and the stability of machine-learning models \cite{MaxAbs}.
Thus, the normalized global error of the type A is given by,
\begin{equation} 
    \begin{split}
    \min_{\alpha}          \quad &Q_{L\_A}(\alpha)=|\frac{L_{err}(f_{exp}, \alpha)}{\mathbf{L_{err\_max}}}|, or\\
                                 &Q_{\phi\_A}(\alpha)=|\frac{\phi_{err}(f_{exp}, \alpha)}{\mathbf{\phi_{err\_max}}}|\\
    \text{subject to}      \quad &\alpha \in [0.5, 1] \\
    \end{split}
    \label{eq_typeA_global_err}
\end{equation} 
where $|\mathbf{L_{err\_max}}|$ is the absolute maximum magnitude error at $f_{exp}$, 
$|\mathbf{\phi_{err\_max}}|$ is the absolute maximum phase error at $f_{exp}$.

The normalized global error of the type B is given by,
\begin{equation} 
    \begin{split}
    \min_{\alpha}         \quad &Q_{L\_B}(\alpha) = \frac{\sqrt{\sum_{i=1}^{N}\mathbf{K_L}[i]\cdot[L_{err}(\mathbf{f_{exp}}[i], \alpha)]^2}}{|\mathbf{L_{err\_max}}|}, or \\
                                &Q_{\phi\_B}(\alpha) = \frac{\sqrt{\sum_{i=1}^{N}\mathbf{K_\phi}[i]\cdot[\phi_{err}(\mathbf{f_{exp}}[i], \alpha)]^2}}{|\mathbf{\phi_{err\_max}}|}\\
    \text{subject to}     \quad &\mathbf{f_{exp}}[i] \in \{\mathbf{f_{exp}}[1], \mathbf{f_{exp}}[2],..., \mathbf{f_{exp}}[N]\} \\
                                &\alpha \in [0.5, 1] \\
    \end{split}
    \label{eq_typeB_global_err}
\end{equation} 
where $\mathbf{K_L}$ and $\mathbf{K_\phi}$ are the magnitude and phase weight of the $i$-th frequency point, respectively.

The normalized global error of the type C is given by,
\begin{equation} 
    \begin{split}
    \min_{\alpha}        \quad &Q_{L\_C}(\alpha) = \frac{\int_{f_{start}}^{f_{end}}|\frac{L_{err}(f_{exp}, \alpha)}{\mathbf{L_{err\_max}}}|\cdot df}{f_{end}-f_{start}}, or \\
                               &Q_{\phi\_C}(\alpha) = \frac{\int_{f_{start}}^{f_{end}}|\frac{\phi_{err}(f_{exp}, \alpha)}{\mathbf{\phi_{err\_max}}}|\cdot df}{f_{end}-f_{start}}\\
    \text{subject to}    \quad &f\in[f_{start}, f_{end}] \\
                               &\alpha \in [0.5, 1] \\
    \end{split}
    \label{eq_typeC_global_err}
\end{equation} 
where $[f_{start}, f_{end}]$ is the expected frequency interval. 

\textbf{Step 5: Computation. }
This step is to calculate the parameter $\alpha_{opt}$ through numerical computation.
To do well, the mathematical tool is important, such as Python, Mathcad, Matlab, etc.
In this paper, we use the Mathcad (V15) as the computing tool, 
and the source file is available in \url{https://github.com/ShaneRun/GBT}.

\subsection{Application Case: LPF}\label{sub2sec4}
\begin{figure}[h]  
  \centering
  \includegraphics[width=0.8\linewidth]{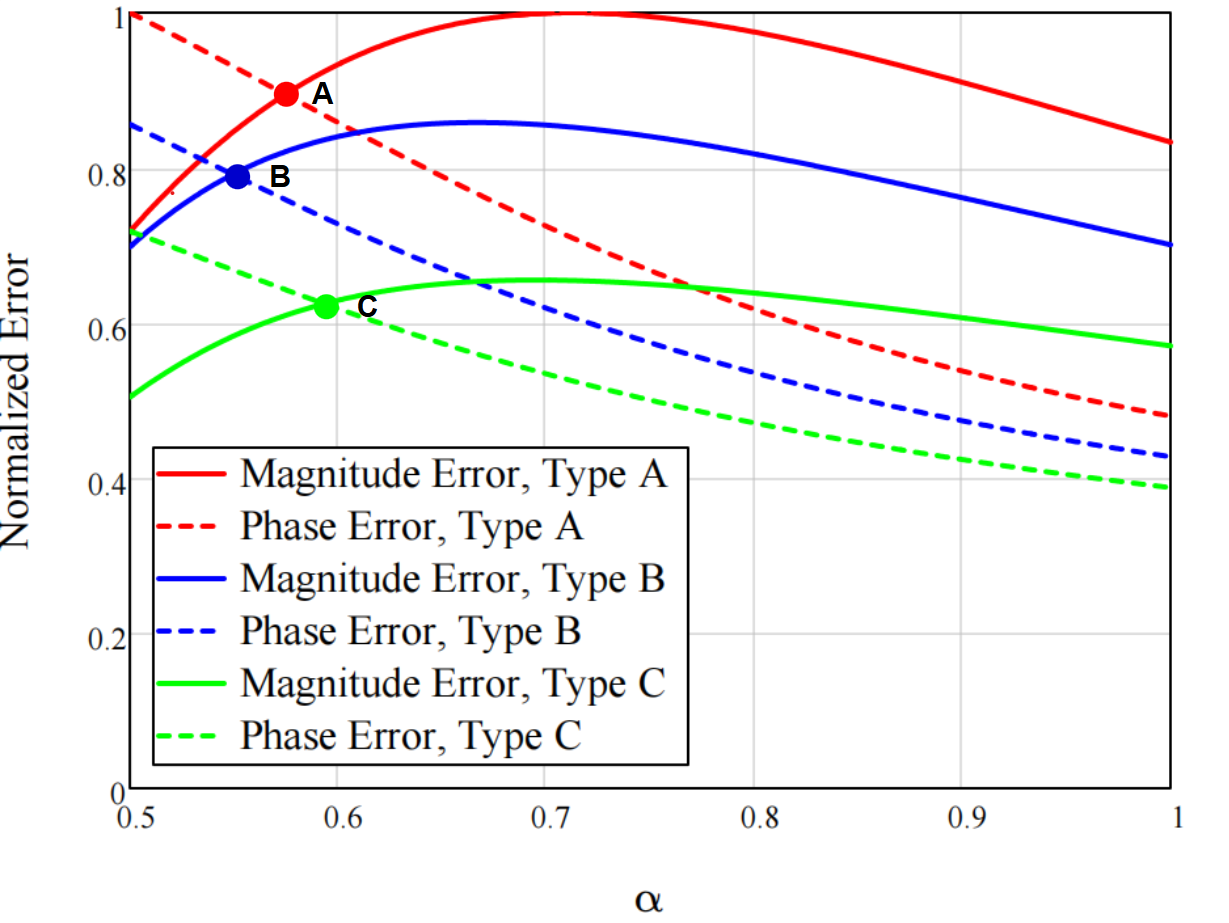}
  \caption{Normalized Global Error vs. $\alpha$}
  \label{fig_norm_err_vs_alpha}
\end{figure}

In this section, 
we apply the optimal design method to discretize an analog LPF covering all application scenarios: type A, type B and type C.
The equivalent resistance and capacitance of the analog LPF are 7.5 $k\Omega$ and 4.4 nF, respectively, 
resulting in a cutoff frequency ($f_c$) of 4.823 kHz.
For Type A scenario, the target frequency point for evaluation is set to 75\% of $f_c$.
For Type B scenario, we refer to the ``CEC (California Energy Commission) Efficiency'' weights as illustrated in Table~\ref{tab:LPF_typeB_weights}.
For Type C scenario, the frequency sweep range is defined from $f_{start}=10\%f_c$ to $f_{end}=f_c$.
The normalized global error of magnitude and phase are shown in Fig.~\ref{fig_norm_err_vs_alpha}, 
and the optimal design results are summarized in Table~\ref{tab:LPF_design_results}.
The trade-off design points corresponding to each application type are marked in Fig.~\ref{fig_norm_err_vs_alpha} as points A, B, and C, respectively.

\begin{table}[!t] 
\caption{Weighting Factor for Discrete LPF(Type B)\label{tab:LPF_typeB_weights}}
\centering
\begin{tabular}{|c|c|c|}
\hline
\textbf{$f_{exp}$} & \textbf{Weighting Factor $K_L or K_\phi$}   \\
\hline
10\%$f_c$        &  0.04 \\
20\%$f_c$        &  0.05 \\
30\%$f_c$        &  0.12 \\
50\%$f_c$        &  0.21 \\
75\%$f_c$        &  0.53 \\
100\%$f_c$       &  0.05 \\
\hline
\end{tabular}
\end{table}

\begin{table}[!t] 
\caption{Optimal Design Results of the Shape Factor $\alpha$\label{tab:LPF_design_results}}
\centering
\begin{tabular}{|p{1.5cm}|p{1.8cm}|p{1.8cm}|p{1.8cm}|}
\hline
\textbf{Scenario Type}    & \textbf{Magnitude First}  & \textbf{Trade-off}$^e$  & \textbf{Phase First} \\
\hline
Type A $^a$               &  0.5 (0.718) $^d$         & 0.575 (0.895)      & 1.0 (0.48)           \\
Type B $^b$               &  0.5 (0.698)              & 0.549 (0.791)      & 1.0 (0.427)          \\
Type C $^c$               &  0.5 (0.504)              & 0.593 (0.625)      & 1.0 (0.388)          \\
\hline
\multicolumn{4}{l}{$^a$ $f_{exp}=$75\%$f_c$} \\
\multicolumn{4}{l}{$^b$ CEC Weights} \\
\multicolumn{4}{l}{$^c$ 10\%-100\% $f_c$} \\
\multicolumn{4}{l}{$^d$ The value inside "()" is the normalized global error } \\
\multicolumn{4}{l}{$^e$ Trade-off design when the normalized errors are equivalent } \\
\end{tabular}
\end{table}

\section{Experimental Evaluation}\label{sec5}
\begin{figure}[h]  
  \centering
  \includegraphics[width=1.0\linewidth]{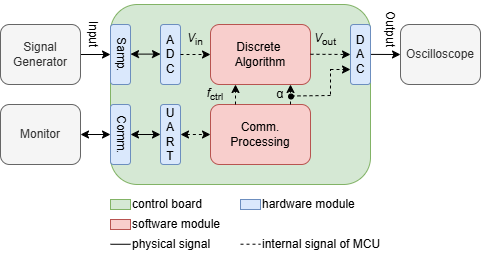}
  \caption{Block diagram of the discrete LPF. Samp.: Sampling, Comm.: Communication, ADC: A/D Conversion, DAC: D/A Conversion.}  
  \label{fig_exp_diagram}
\end{figure}

\begin{figure}[h]  
  \centering
  \includegraphics[width=0.8\linewidth]{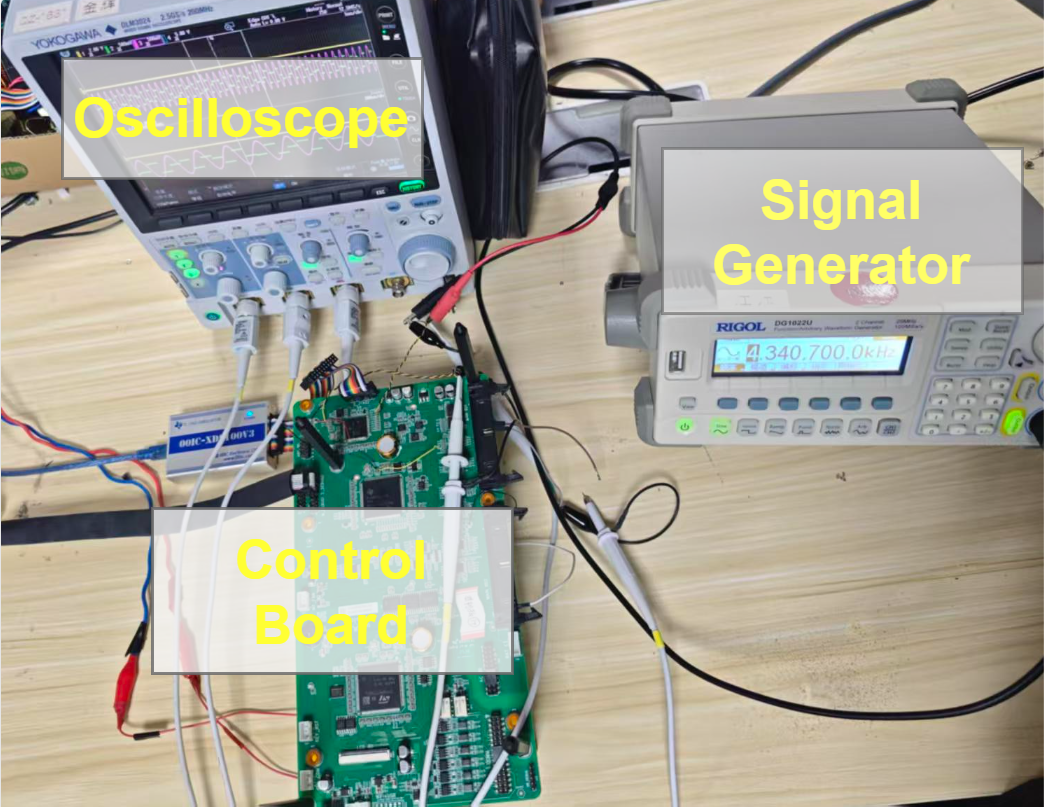}
  \caption{Photo of experimental setup.}
  \label{fig_exp_setup}
\end{figure}

To validate the effectiveness of the proposed discrete frequency response and the optimal design method,
a discrete LPF is built as shown in Fig.~\ref{fig_exp_diagram}.
The algorithms are deployed on a control board equipped with a TMS320F28P65 microcontroller.
A high-precision signal generator, DG1022U, provides the input signal
A laptop serves as a monitoring terminal and communicates with the control board via RS485 to configure parameters 
such as the sampling frequency ($f_{samp}$) and the shape factor ($\alpha$).
The "Discrete Algorithm" module implements the discrete LPF in the form of a difference equation.
As expressed in equation ~\eqref{eq_discrete_plant}, the discrete transfer function of the LPF is defined as follows:
\begin{equation}
  \begin{split}
  G_{LPF\_disc}(z,\alpha) &= \frac{V_{out}(z)}{V_{in}(z)}\\
                          &= \frac{\alpha w_c T + [(1-\alpha) w_c T]z^{-1}}{1+\alpha w_c T + [(1-\alpha) w_c T - 1]z^{-1}}    
  \end{split}
  \label{eq_lpf_z_form}
\end{equation}
Therefore, the difference equation of the discrete LPF is given by,
\begin{equation}
  \begin{split}
  V_{out}(n) = &V_{out}(n-1) + \frac{\alpha w_c T}{1+\alpha w_c T}\cdot[V_{in}(n)-V_{in}(n-1)] +\\
                &\frac{w_c T}{1+\alpha w_c T}\cdot[V_{in}(n-1)-V_{out}(n-1)] + \\
  \end{split}  
  \label{eq_lpf_diff_form}
\end{equation}
where $V_{x}(n)$ and $V_{x}(n-1)$ (x="in" or "out") are the computed results of current and last sampling period, respectively.
The relevant variables and signals are sent to the D/A converter to be observed in an oscilloscope.
The experiment setup is shown in Fig~\ref{fig_exp_setup}.

\subsection{Verification of the Discrete Frequency Response}\label{sub1sec5}
\begin{figure}[h]  
  \centering
  \includegraphics[width=1.0\linewidth]{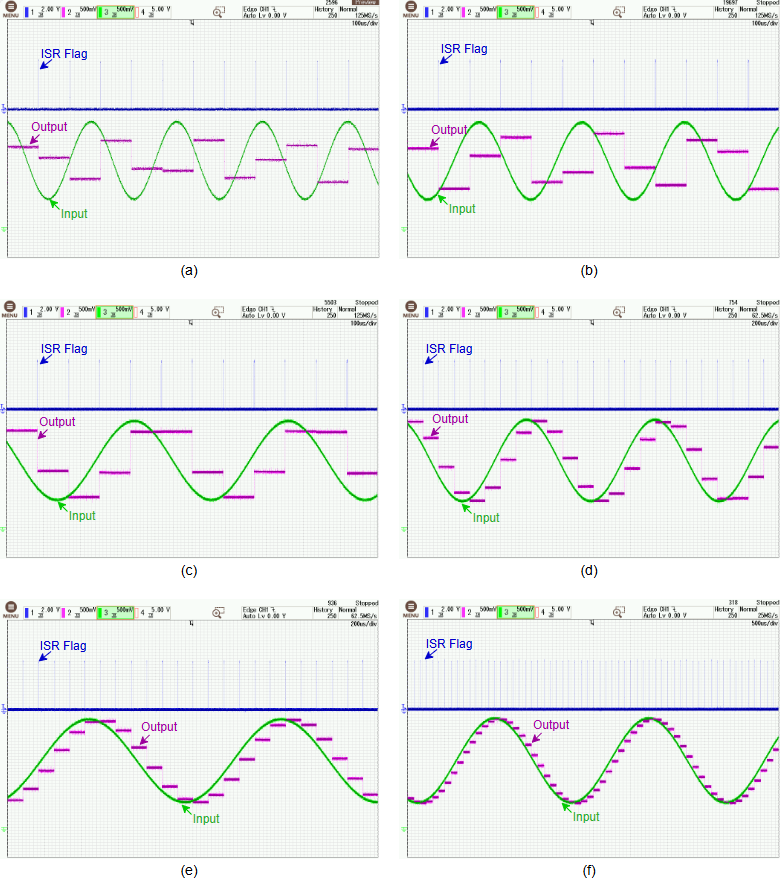}
  \caption{Experimental waveforms with different input frequency. CH1, ISR (Interrupt Service Routine) flag, 2 V/div, CH2, output voltage, 0.5 V/div; CH3, input voltage, 0.5 V/div. 
(a) $f=100\%f_c$, time, 100 us/div. 
(b) $f=75\%f_c$, time, 100 us/div.
(c) $f=50\%f_c$, time, 100 us/div. 
(d) $f=30\%f_c$, time, 200 us/div.
(e) $f=20\%f_c$, time, 200 us/div.
(f) $f=10\%f_c$, time, 500 us/div.
}\label{fig_exp_waveform_1}
\end{figure}

\begin{figure}[h]  
  \centering
  \includegraphics[width=0.8\linewidth]{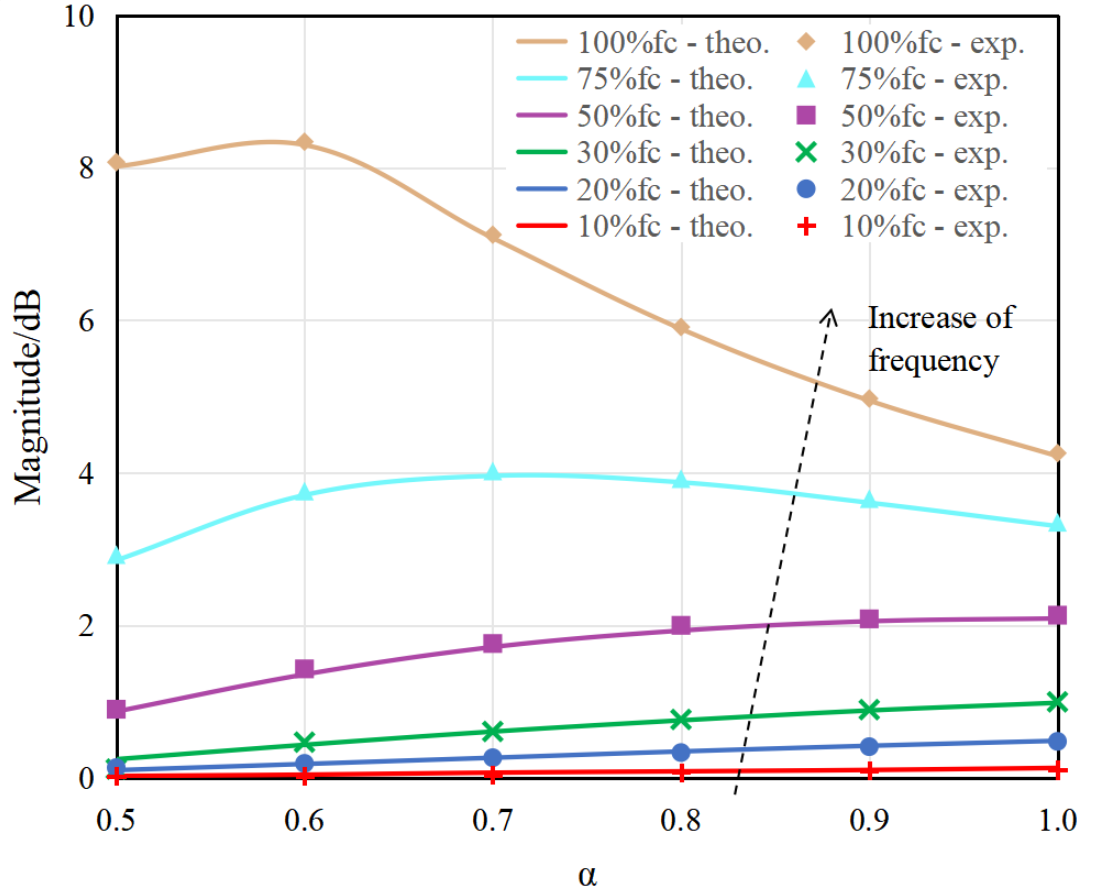}
  \caption{Comparison of magnitude between the theoretical calculations and the experimental results. theo.: theoretical, exp.: experimental.}
  \label{fig_mag_vs_f}
\end{figure}

\begin{figure}[h]  
  \centering
  \includegraphics[width=0.8\linewidth]{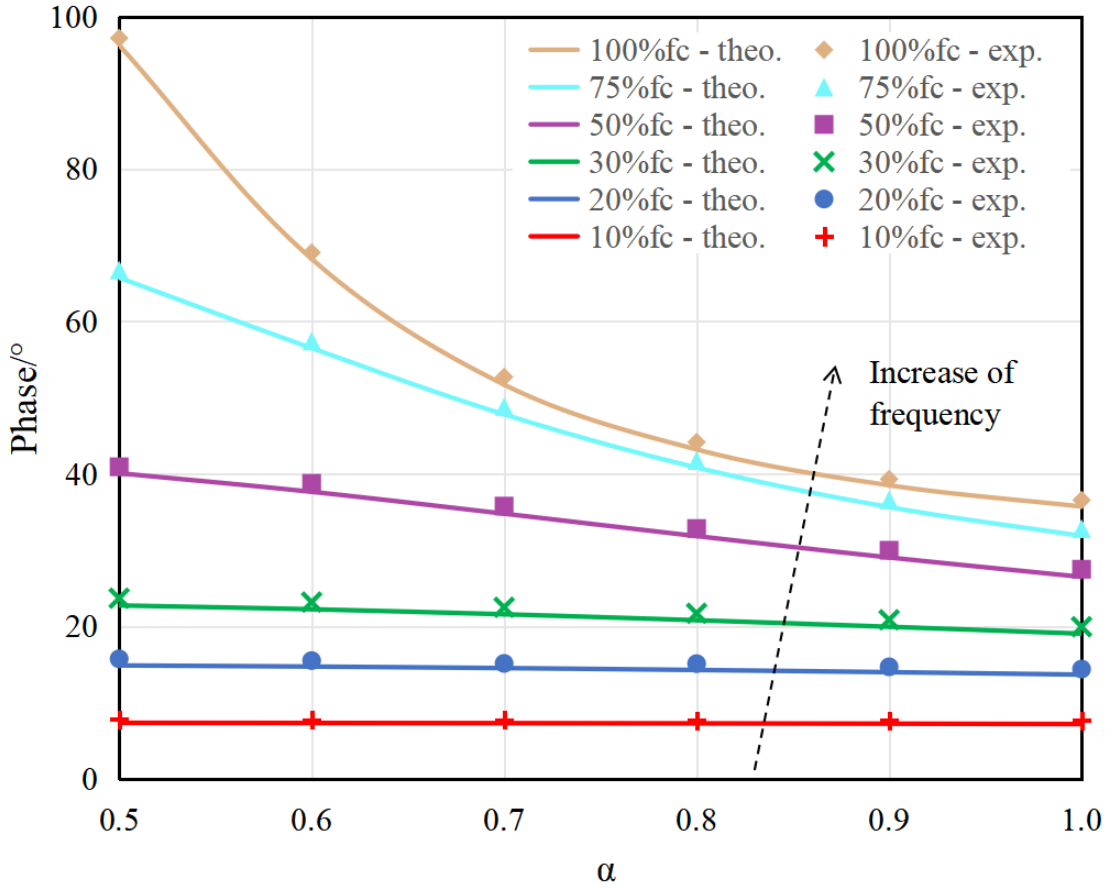}
  \caption{Comparison of phase between theoretical calculations and experimental results. theo.: theoretical, exp.: experimental.}
  \label{fig_ph_vs_f}
\end{figure}

To validate the proposed discrete frequency response as expressed in \eqref{eq_lpf_discrete}, 
the testing frequency set is \{$10\%f_c, 20\%f_c, 30\%f_c, 50\%f_c, 75\%f_c, 100\%f_c$\}, 
and the corresponding shape factor $\alpha$ all starts at 0.5 and increases by 0.1 for each testing frequency. 
The experimental waveforms of the input and output voltage are shown in Fig.~\ref{fig_exp_waveform_1}.
In this case, $\alpha$ is fixed at 0.5, 
and the testing frequency is adjusted based on the testing frequency set.
As shown in Fig.~\ref{fig_mag_vs_f}, 
the magnitude errors between the theoretical calculations and the experimental results 
for different testing frequencies are quite small (error rate $\leq$ 5\%),
which verifies the effectiveness of the proposed discrete frequency response.
The detailed magnitude error data are listed in Table~\ref{tab:LPF_exp_mag_err_rate}.
The phase error curve and data are shown in Fig.~\ref{fig_ph_vs_f} and Table~\ref{tab:LPF_exp_phase_err_rate}, respectively.

\begin{table}[h] 
\caption{Error Rate of Magnitude\label{tab:LPF_exp_mag_err_rate}}
\centering
\begin{tabular}{|p{2.8cm}|p{1.5cm}|p{1.5cm}|p{1.2cm}|}
\hline
\textbf{Test Cases}    & \textbf{Theoretical Error/dB}  & \textbf{Experimental Error/dB}  & \textbf{Error Rate$^a$/\%} \\
\hline
$\alpha=0.5, f=75\%f_c$   &  3.30         &  3.35   &  1.53      \\
$\alpha=0.6, f=75\%f_c$   &  3.61         &  3.66   &  1.48      \\
$\alpha=0.7, f=75\%f_c$   &  3.88         &  3.92   &  1.24      \\
$\alpha=0.8, f=75\%f_c$   &  3.96         &  4.03   &  1.60      \\
$\alpha=0.9, f=75\%f_c$   &  3.71         &  3.76   &  1.43      \\
$\alpha=1.0, f=75\%f_c$   &  2.85         &  2.93   &  2.63      \\
$\alpha=0.5, f=f_c$       &  4.22         &  4.26   &  0.92      \\
$\alpha=0.6, f=f_c$       &  4.95         &  4.97   &  0.52      \\
$\alpha=0.7, f=f_c$       &  5.89         &  5.91   &  0.41      \\
$\alpha=0.8, f=f_c$       &  7.08         &  7.12   &  0.61      \\
$\alpha=0.9, f=f_c$       &  8.30         &  8.34   &  0.51      \\
$\alpha=1.0, f=f_c$       &  8.02         &  8.07   &  0.72      \\
\hline
\multicolumn{4}{l}{$^a$ Error Rate=(Experimental Error - Theoretical Error)/Theoretical Error} \\
\end{tabular}
\end{table}

\begin{table}[h] 
\caption{Error Rate of Phase\label{tab:LPF_exp_phase_err_rate}}
\centering
\begin{tabular}{|p{2.8cm}|p{1.5cm}|p{1.5cm}|p{1.2cm}|}
\hline
\textbf{Test Cases}    & \textbf{Theoretical Error/$^\circ$}  & \textbf{Experimental Error/$^\circ$}  & \textbf{Error Rate$^a$/\%} \\
\hline
$\alpha=0.5, f=75\%f_c$   &  31.25         & 32.83     &  5.06   \\
$\alpha=0.6, f=75\%f_c$   &  35.04         & 36.60     &  4.45   \\
$\alpha=0.7, f=75\%f_c$   &  40.23         & 41.76     &  3.80   \\
$\alpha=0.8, f=75\%f_c$   &  47.20         & 48.84     &  3.47   \\
$\alpha=0.9, f=75\%f_c$   &  55.90         & 57.45     &  2.77   \\
$\alpha=1.0, f=75\%f_c$   &  65.13         & 66.75     &  2.49   \\
$\alpha=0.5, f=f_c$       &  34.90         & 36.57     &  4.79   \\
$\alpha=0.6, f=f_c$       &  37.70         & 39.32     &  4.30   \\
$\alpha=0.7, f=f_c$       &  42.40         & 44.18     &  4.20   \\
$\alpha=0.8, f=f_c$       &  50.90         & 52.73     &  3.60   \\
$\alpha=0.9, f=f_c$       &  67.41         & 69.05  	 &  2.43   \\
$\alpha=1.0, f=f_c$       &  95.45         & 97.19     &  1.82   \\
\hline
\multicolumn{4}{l}{$^a$ Error Rate=(Experimental Error - Theoretical Error)/Theoretical Error} \\
\end{tabular}
\end{table}

\begin{figure}[h]  
  \centering
  \includegraphics[width=0.8\linewidth]{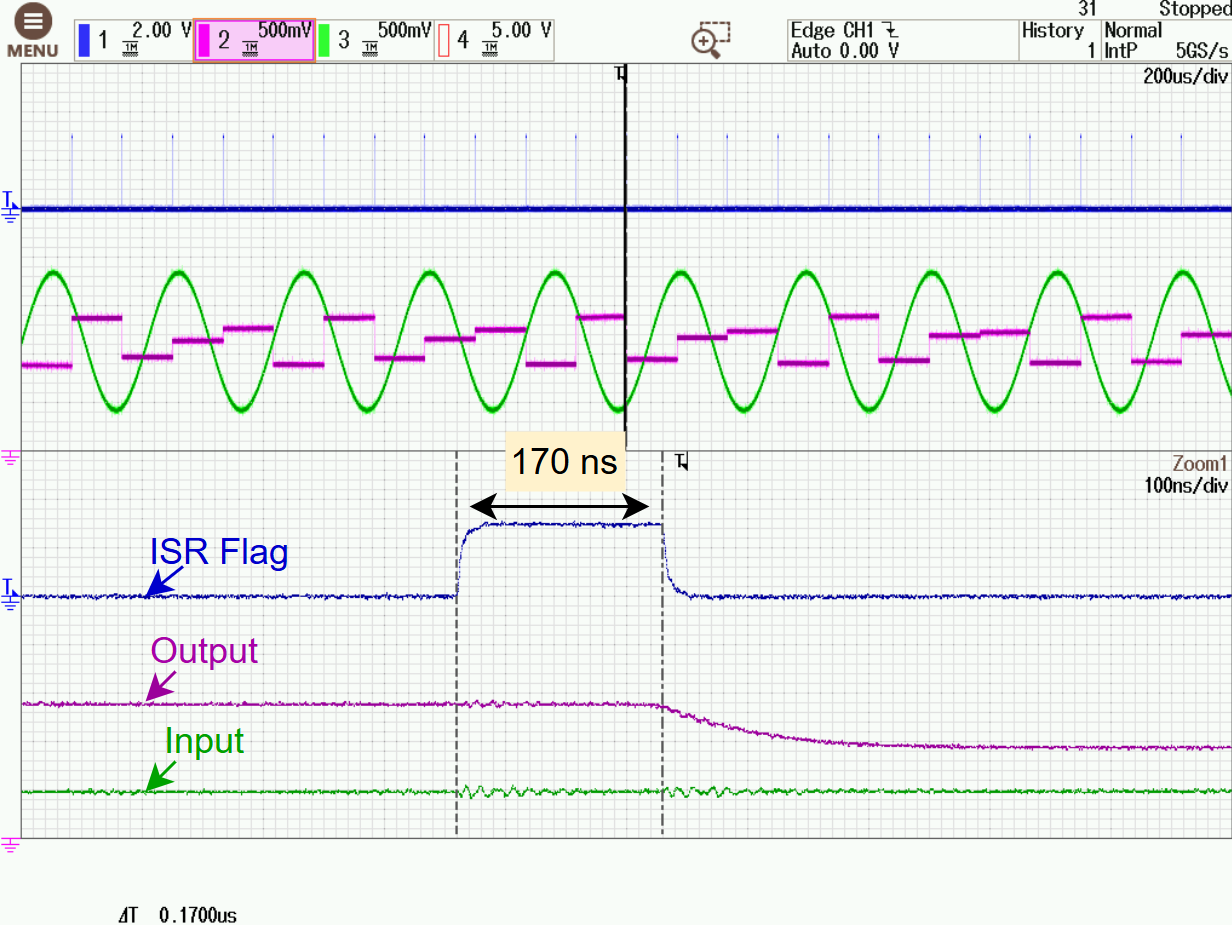}
  \caption{Experimental waveform of the time delay ($T_{cmpt}+T_{dac}$).}
  \label{fig_waveform_delay}
\end{figure}

By comparing Fig.~\ref{fig_mag_vs_f} and Fig.~\ref{fig_ph_vs_f}, 
we observe that the phase error is significantly larger than the magnitude error.
This is mainly due to the time delay, $T_{delay}$, for digital signal processing,
which consists of the sample-and-hold time of ADC, $T_{adc}$, 
the computation time of CPU, $T_{cmpt}$, 
and the digital-to-analog conversion time of DAC, $T_{dac}$.
In this article, $T_{adc}$ is set to 300 ns.
We use the assembly code to estimate $T_{cmpt}$, 
and the execution cycles for all instructs are estimated to be twenty, including 
eight times of ``MOV32''(1 cycle), four times of ``ADDF32''(2 cycles),
four times of ``MPYF32''(2 cycles), four times of ``SUBF32''(2 cycles), and zero times of ``DIVF32''(5 cycles).
Therefore, the theoretical value of $T_{cmpt}$ is estimated to be 100 ns
by multiplying the ``total cycles'' by the ``cycle time''(5 ns @ 200 MHz).
We use the immediate loading scheme, 
and the theoretical value of $T_{dac}$ is about 70 ns via test.
The measured value of ($T_{cmpt}+T_{dac}$) is 170 ns as shown in Fig~\ref{fig_waveform_delay}.
Therefore, $T_{delay}$ is 470ns.

Time delay does not affect the magnitude response, so we focus exclusively on its impact on phase.
Using delay compensation, we calibrate the experimental phase error by subtracting the measured phase error
by subtracting the measured phase error (a negative value)
from the additional phase lag (also negative) introduced by the time delay.
A comparison of the phase errors with and without delay compensation shows that 
the phase error is significantly reduced after compensation, by approximately 45\%.

\begin{figure}[h]  
  \centering
  \includegraphics[width=0.8\linewidth]{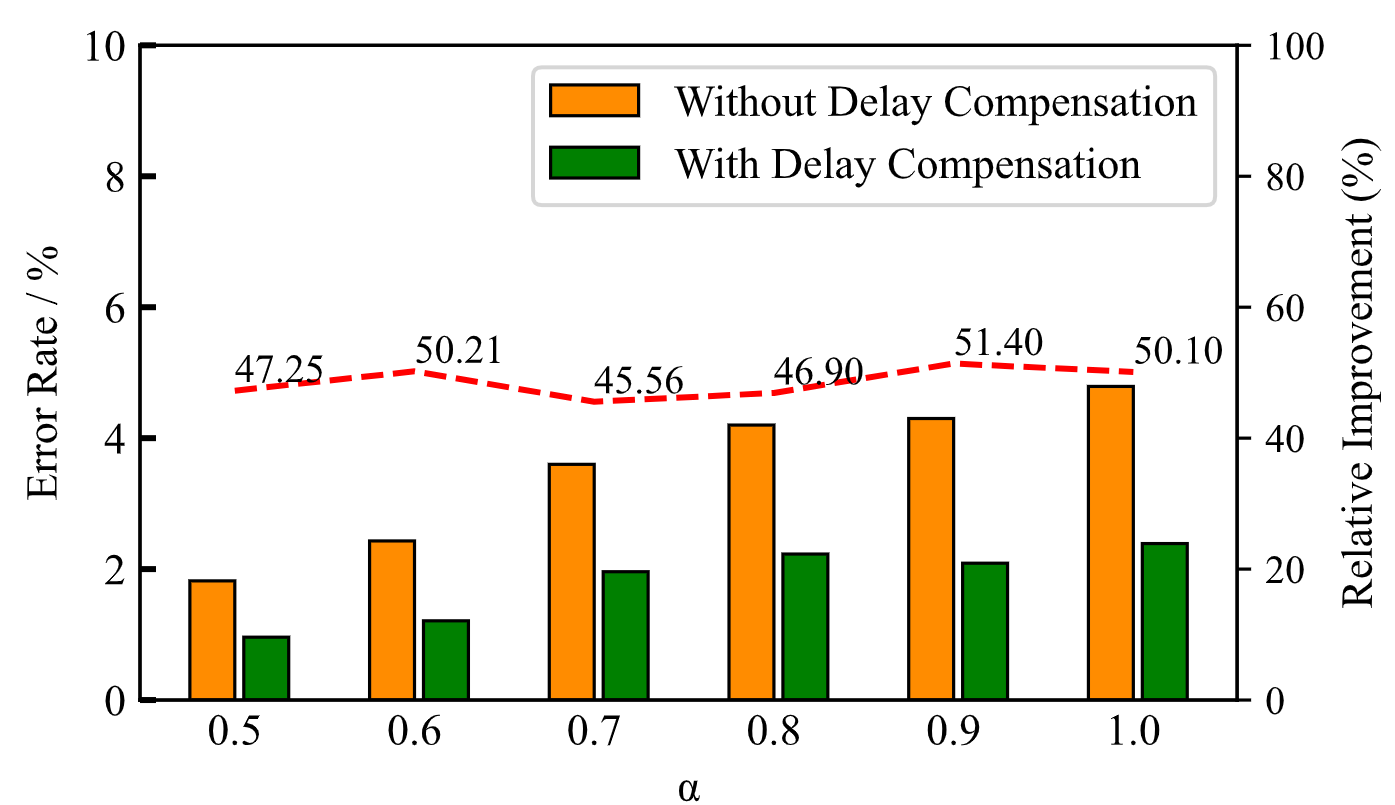}
  \caption{Comparison of the phase error between with and without delay compensation.}
  \label{fig_delay_effect}
\end{figure}

\subsection{Verification of Optimal Design Method}\label{sub2sec5}
\begin{figure}[h]  
  \centering
  \includegraphics[width=0.8\linewidth]{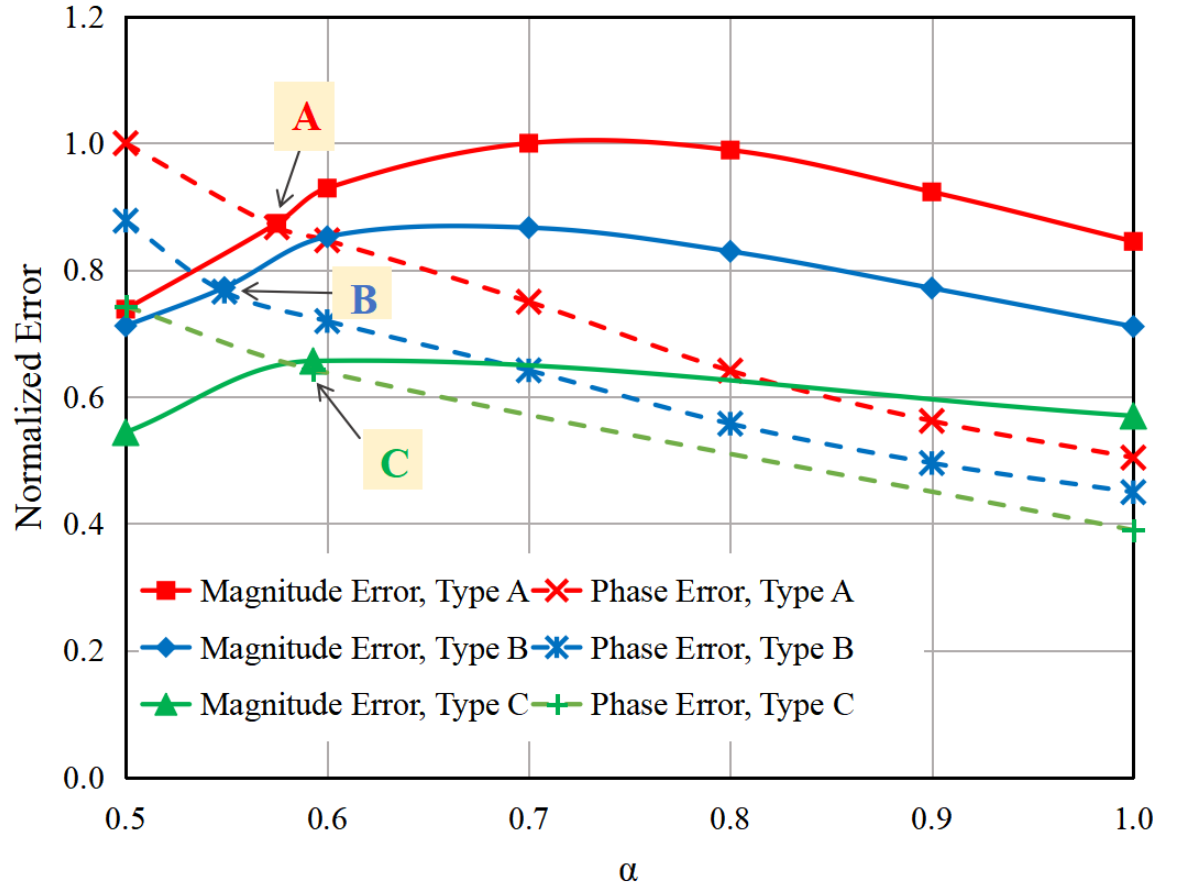}
  \caption{Experimental Global Error vs. $\alpha$}
  \label{fig_exp_err_vs_alpha}
\end{figure}

To validate the proposed optimal design method, 
we adopt the testing scenarios and operating frequency settings presented in Section\ref{sub2sec4}.
The experimental normalized error curve for both magnitude and phase are shown in Fig.~\ref{fig_exp_err_vs_alpha}, 
with corresponding error data provided in Table~\ref{tab:LPF_exp_check_design_results}.
For different application scenarios, 
the experimental trade-off points are marked as A, B, and C in Fig.~\ref{fig_exp_err_vs_alpha}, respectively.
Both the curves and the quantitative data indicate a low error level, 
which confirm the effectiveness of the proposed method.

\begin{figure}[h]  
  \centering
  \includegraphics[width=0.8\linewidth]{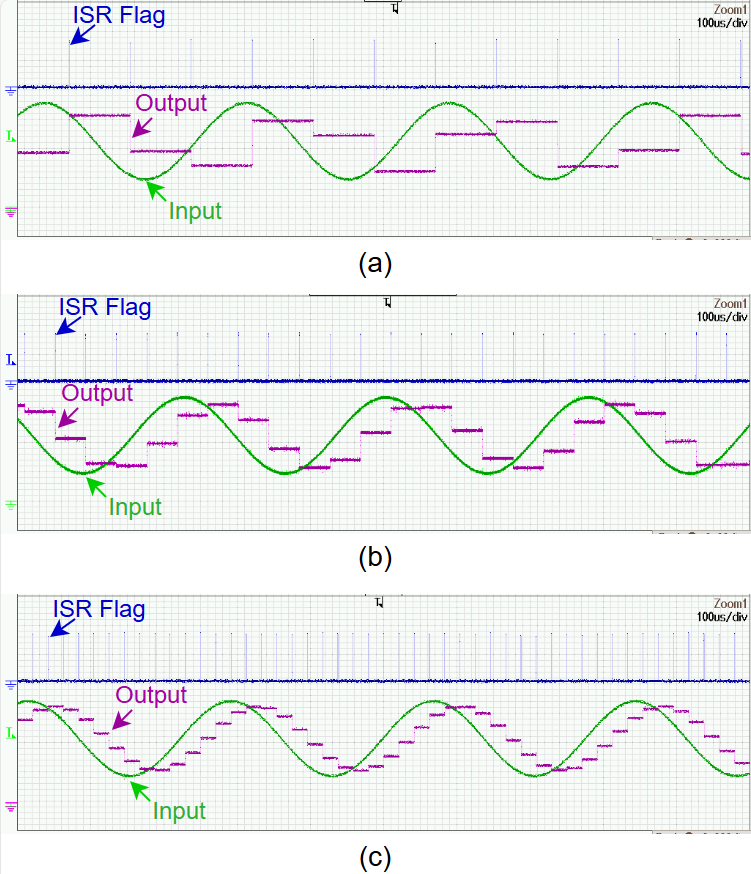}
  \caption{Experimental waveforms with different sampling frequency. CH1, ISR flag, 2 V/div, CH2, output voltage, 0.5 V/div; CH3, input voltage, 0.5 V/div; time, 100 us/div. 
  (a) $f_{samp}$=12 kHz. 
  (b) $f_{samp}$=24 kHz. 
  (c) $f_{samp}$=48 kHz. 
}\label{fig_exp_waveform_2}
\end{figure}

\begin{figure}[h]  
  \centering
  \subfloat[]{\includegraphics[width=0.5\linewidth]{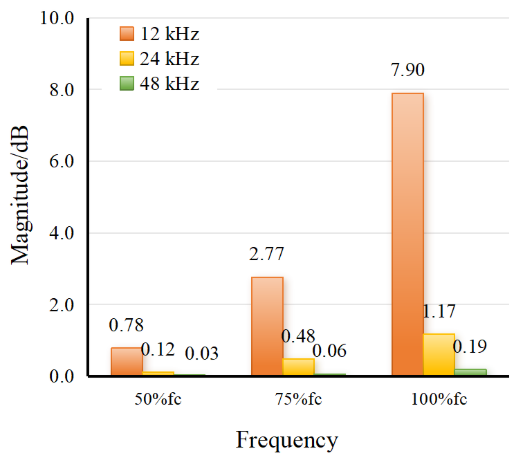}\label{(a)}}
  \hfil
  \subfloat[]{\includegraphics[width=0.5\linewidth]{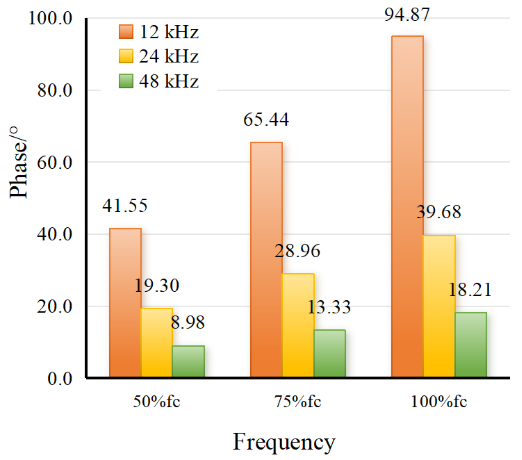}\label{(b)}}
  \caption{Comparison of magnitude and phase error under different sampling frequency.}
  \label{fig_err_vs_fctrl}
\end{figure}

\begin{table}[h] 
\caption{Experimental Verification of Optimal Design Results\label{tab:LPF_exp_check_design_results}}
\centering
\begin{tabular}{|p{1.8cm}|p{0.8cm}|p{1.5cm}|p{1.5cm}|p{1.2cm}|}
\hline
\textbf{Scenario Type}    &\textbf{$\alpha_{opt}$}& \textbf{Theoretical Error}  & \textbf{Experimental Error}  & \textbf{Error Rate/\% $^a$} \\
\hline
A, Mag. First      & 0.5    &  0.718         &  0.738  &  2.82    \\
A, Trade-off       & 0.575  &  0.895         &  0.874  &  -2.39   \\
A, Phase First     & 1.0    &  0.480         &  0.504  &  5.02    \\
B, Mag. First      & 0.5    &  0.698         &  0.712  &  2.00    \\
B, Trade-off       & 0.549  &  0.791         &  0.769  &  -2.78   \\
B, Phase First     & 1.0    &  0.427         &  0.450  &  5.29    \\
C, Mag. First      & 0.5    &  0.504         &  0.544  &  7.93    \\
C, Trade-off       & 0.593  &  0.625         &  0.650  &  3.97    \\
C, Phase First     & 1.0    &  0.388         &  0.390  &  0.44    \\
\hline
\multicolumn{5}{l}{$^a$ Error Rate=(Experimental Error - Theoretical Error)/Theoretical Error} \\
\end{tabular}
\end{table}

In summary, experimental results verify the effectiveness of the proposed optimal design method.
However, a significant limitation is observed: discretization-induced distortion becomes severe 
as the operating frequency approaches the Nyquist frequency. 
To alleviate this issue, a practical solution is to increase the sampling frequency. 
In this study, the magnitude and phase error are measured when the sampling frequency is raised from 12 kHz to 48 kHz,
while $\alpha$ is fixed at 0.5. 
Fig~\ref{fig_exp_waveform_2} compares the experimental waveforms under different sampling frequencies,
and Fig.~\ref{fig_err_vs_fctrl} provides the corresponding error comparison.
As shown in Fig~\ref{fig_exp_waveform_2}, the output voltage becomes more sinusoidal with higher sampling frequency.
As shown in Fig.~\ref{fig_err_vs_fctrl}(a), the magnitude error at $f_c$ is significantly reduced from 7.9 dB to 0.19 dB 
when $f_{samp}$ is raised from 12 kHz to 48 kHz, 
while the phase error is reduced from 94.87$^\circ$ to 18.21$^\circ$ in this case as shown in Fig.~\ref{fig_err_vs_fctrl}(b).
These results clearly demonstrate that a higher sampling frequency substantially reduces distortion in the discrete LPF.

\section{Conclusion}\label{sec6}
Discretization is an essential step in implementing digital systems, and selecting appropriate discretization methods is crucial for performance and accuracy.
In this article, we present an in-depth analysis of the GBT method and demonstrate that the GBT can be served as a unified framework for numerical integration.
The main conclusions of this study are summarized as follows: 
\begin{enumerate}
    \item A novel hexagonal approximation for the GBT is derived by employing a new hexagonal shape 
to approximate the enclosed area of the error function, and we define the parameter $\alpha$ as the shape factor.
Meanwhile, we demonstrate that the physical meaning of the shape factor $\alpha$ is the percentage of the backward rectangular area.
   \item  Two distortion modes are identified for hexagonal approximation with different 
operating frequencies and shape factor: i.e., the magnitude and the phase distortion. 
    \item An optimal design method for the shape factor $\alpha$ is proposed
based on an objective function in form of the normalized magnitude or phase error. 
Comparisons between the theoretical calculations and experimental results verify
its excellent performance for reducing discretization error under different operating frequency conditions. 
\end{enumerate}

\section*{Acknowledgments}
This work was supported by the National Key Research and Development Program of China under Grant 2023YFB2604600.

{\appendix[Review on the Discretization Methods]
We searched IEEEXplore using the keyword "discretization" in "\textit{IEEE Trans. Power Electron.}", "\textit{IEEE Trans. Ind. Inform.}", and "\textit{IEEE Trans. Ind. Electron.}", 
there were a total of 155 results.
After filtering manually, 81 results were found that were related to the topic of discretization methods: \cite{Direct-01} -- \cite{Others-02}.
The distribution graph is shown in Fig.~\ref{fig_method_distribution}, and the summary is presented in Table~\ref{tab_review_result}.
\begin{table}[h] 
\caption{Distribution of Discretization Methods\label{tab_review_result}}
\centering
\begin{tabular}{|c|c|c|c|}
\hline
\textbf{Method}                 & \textbf{Count} & \textbf{Ratio} & \textbf{Literatures}  \\
\hline
Direct Discrete                 & 5              & 6.2\%          & \cite{Direct-01} -- \cite{Direct-05}  \\
Euler                           & 48             & 59.3\%         & \cite{Euler-01} -- \cite{Euler-48}    \\
Modified Euler (e.g., Heun)     & 4              & 4.9\%          & \cite{Heun-01} -- \cite{Heun-04}  \\
Tustin                          & 8              & 9.9\%         & \cite{Tustin-01} -- \cite{Tustin-08}  \\
SOTE                            & 4              & 4.9\%          & \cite{SOTE-01} -- \cite{SOTE-04}  \\
HOTE                            & 7              & 8.6\%          & \cite{HOTE-01} -- \cite{HOTE-RK-03}  \\
Exact                           & 3              & 3.7\%          & \cite{Exact-01} -- \cite{Exact-03}  \\
Others                          & 2              & 2.4\%          & \cite{Others-01} -- \cite{Others-02} \\
\hline
\end{tabular}
\end{table}
}

\newpage

\section{Biography Section}
\begin{IEEEbiography}[{\includegraphics[width=1in,height=1.25in,clip,keepaspectratio]{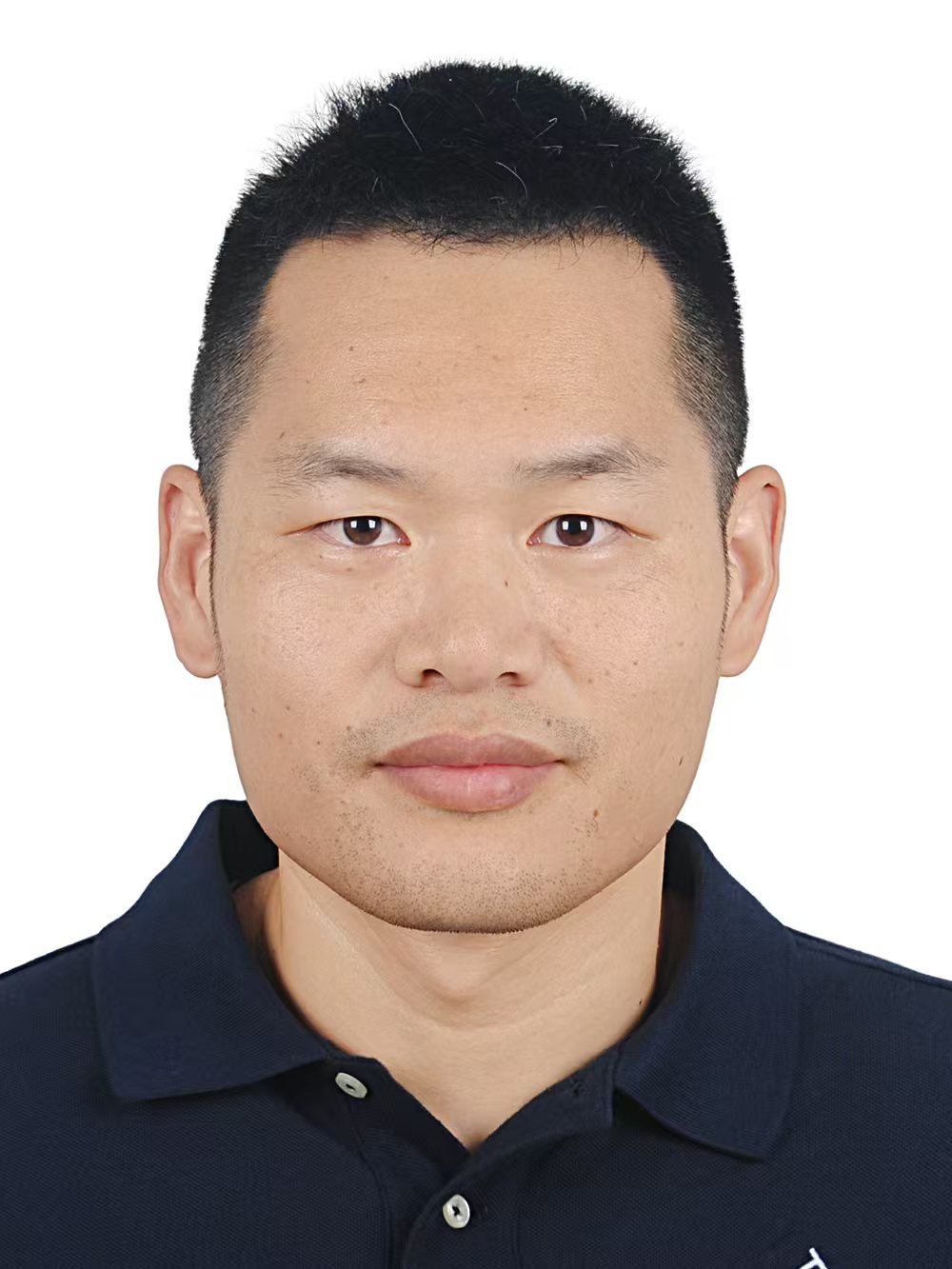}}]{Shen Chen}
(Student Member, IEEE) received the
B.S. degree in electrical engineering from Fuzhou university, Fuzhou, China, 
and the M.S. degree from Zhejiang university, Hangzhou, China, in 2009, and 2012, respectively.
He is currently working toward the Ph.D. degree with Xi'an Jiaotong University, Xi'an, China,
and a software leader with SolaX Power Network Technology (Zhejiang) Co., Ltd., Hangzhou, China.

His research interests include modeling, and control of inverters and system design of microgrids.
\end{IEEEbiography}

\begin{IEEEbiography}[{\includegraphics[width=1in,height=1.25in,clip,keepaspectratio]{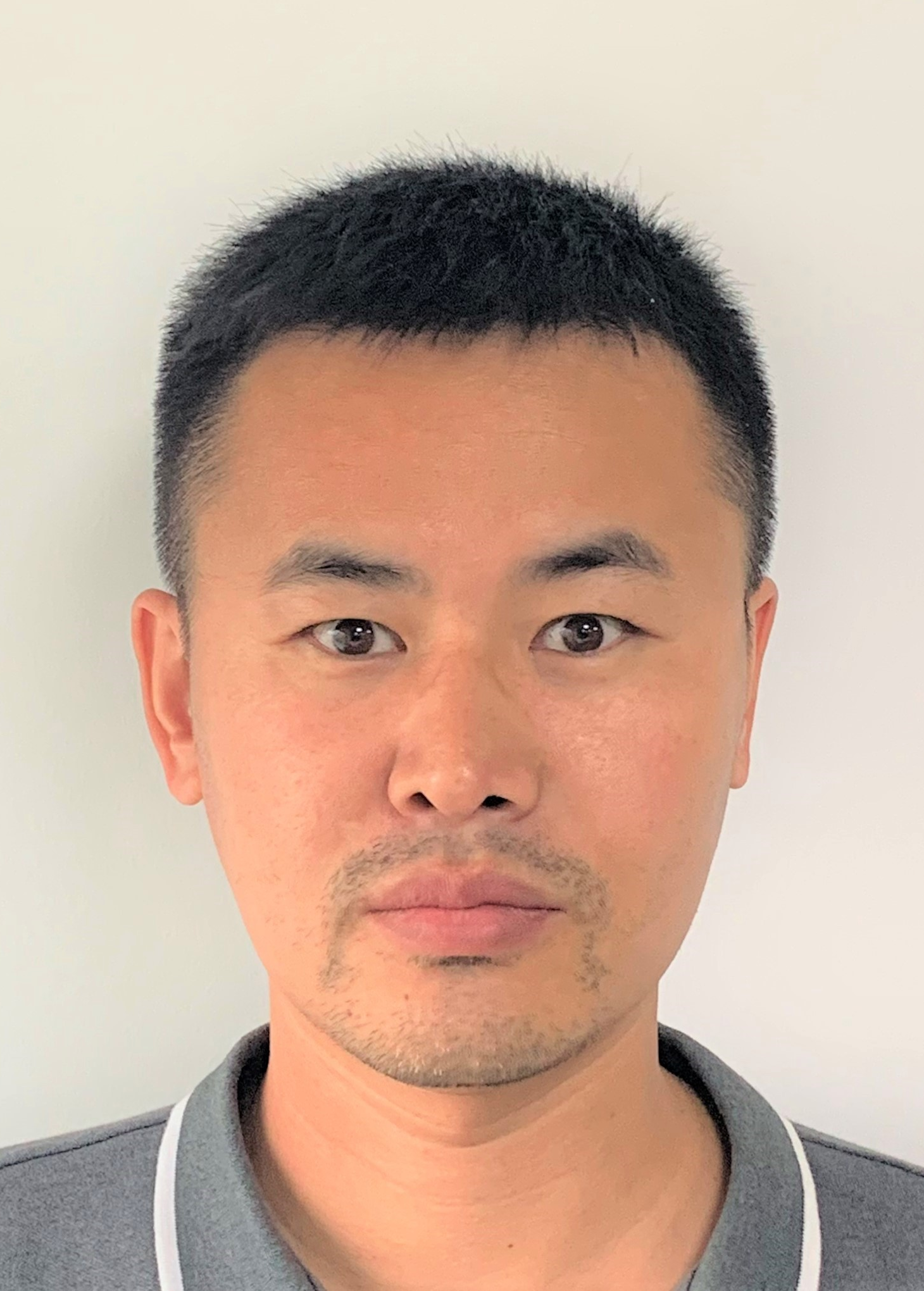}}]{Yanlong Li}
received the B.S. degree in electrical engineering from Harbin University of Science and Technology, Harbin, China, 
and the M.S. degree from Huazhong University of Science and Technology, Wuhan, China, in 2009, and 2011, respectively. 
He is currently a R\&D leader with SolaX Power Network Technology (Zhejiang) Co., Ltd., Hangzhou, China.

His research interests include high-efficiency power conversion in renewable energy systems 
for both residential and commercial applications.
\end{IEEEbiography}

\begin{IEEEbiography}[{\includegraphics[width=1in,height=1.25in,clip,keepaspectratio]{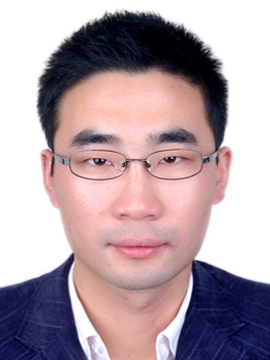}}]{Jiamin Cui} 
received the B.S. and M.S. degrees in electronic engineering from the Ludong University and 
Hangzhou Dianzi University, China, in 2005 and 2008, respectively.  
He is currently a teacher of the College of Electrical engineering, 
Zhejiang University of Water Resources and Electric Power. 

His research interests include communication power supply, intelligent battery charger, 
machine learning in power electronics, etc.
\end{IEEEbiography}

\begin{IEEEbiography}[{\includegraphics[width=1in,height=1.25in,clip,keepaspectratio]{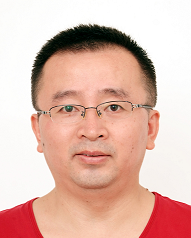}}]{Wei Yao} 
received the M.S degree in Control Theory and Control Engineering from Zhejiang University of Technology 
and the Ph.D. degree in Electric Engineering from Zhejiang University, in 2002 and 2015, respectively. 
He is currently an associate professor with College of Electric Engineering, 
Zhejiang University of Water Resources and Electric Power, Hangzhou, China.

His research interests include motor control, and power electronics, etc.
\end{IEEEbiography}

\begin{IEEEbiography}[{\includegraphics[width=1in,height=1.25in,clip,keepaspectratio]{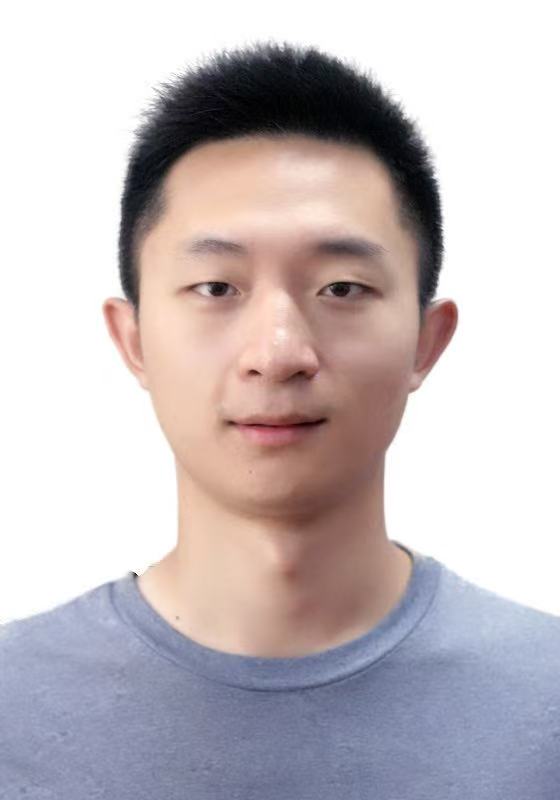}}]{Jisong Wang}
received the B.S. degree in electrical engineering from Yanshan University, Qinhuangdao, China, 
and  the M.S. degree from Naval University of Engineering, Wuhan, China, in 2016, and 2019, respectively.
He is currently a software manager with SolaX Power Network Technology (Zhejiang) Co., Ltd., Hangzhou, China.

His research interests include modeling, and control of inverters.
\end{IEEEbiography}

\begin{IEEEbiography}[{\includegraphics[width=1in,height=1.25in,clip,keepaspectratio]{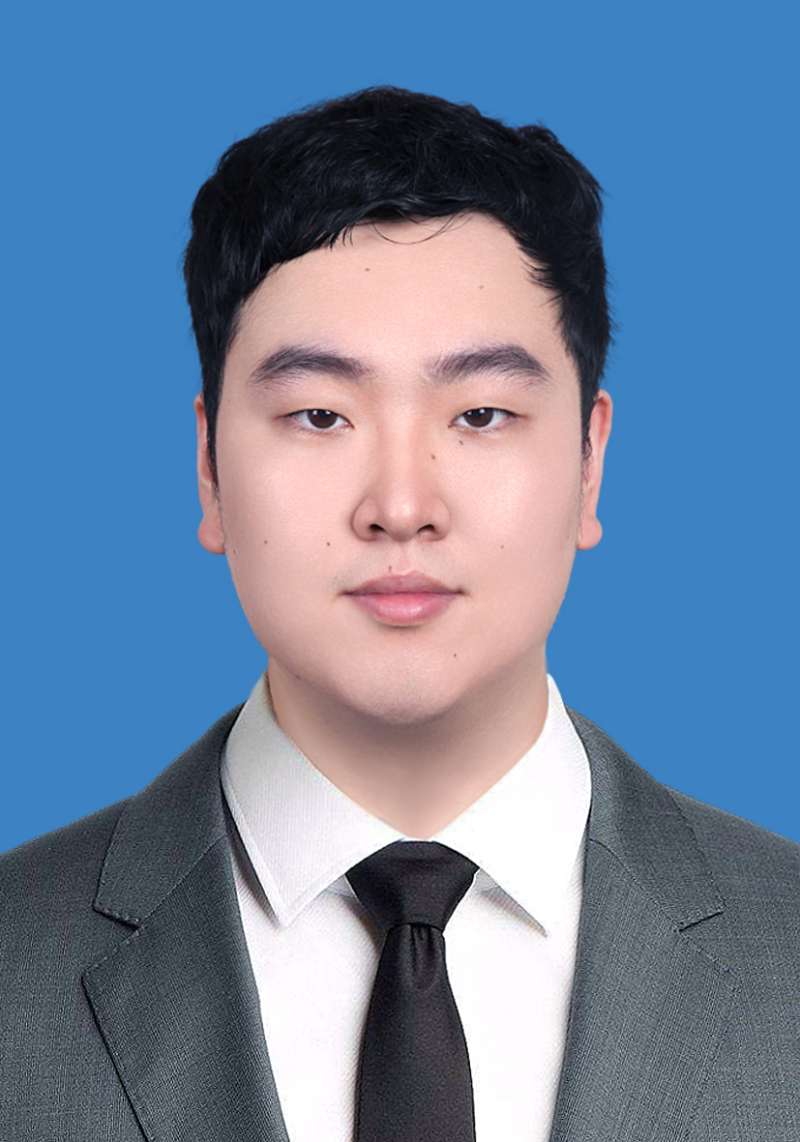}}]{Yixin Tian}
received the B.S. degree in electrical engineering from University of Technology and Education, Tianjin, China, in 2024.
He is currently a software engineer with SolaX Power Network Technology (Zhejiang) Co., Ltd., Hangzhou, China.

His research interests include human-machine interface, and control of inverters.
\end{IEEEbiography}

\begin{IEEEbiography}[{\includegraphics[width=1in,height=1.25in,clip,keepaspectratio]{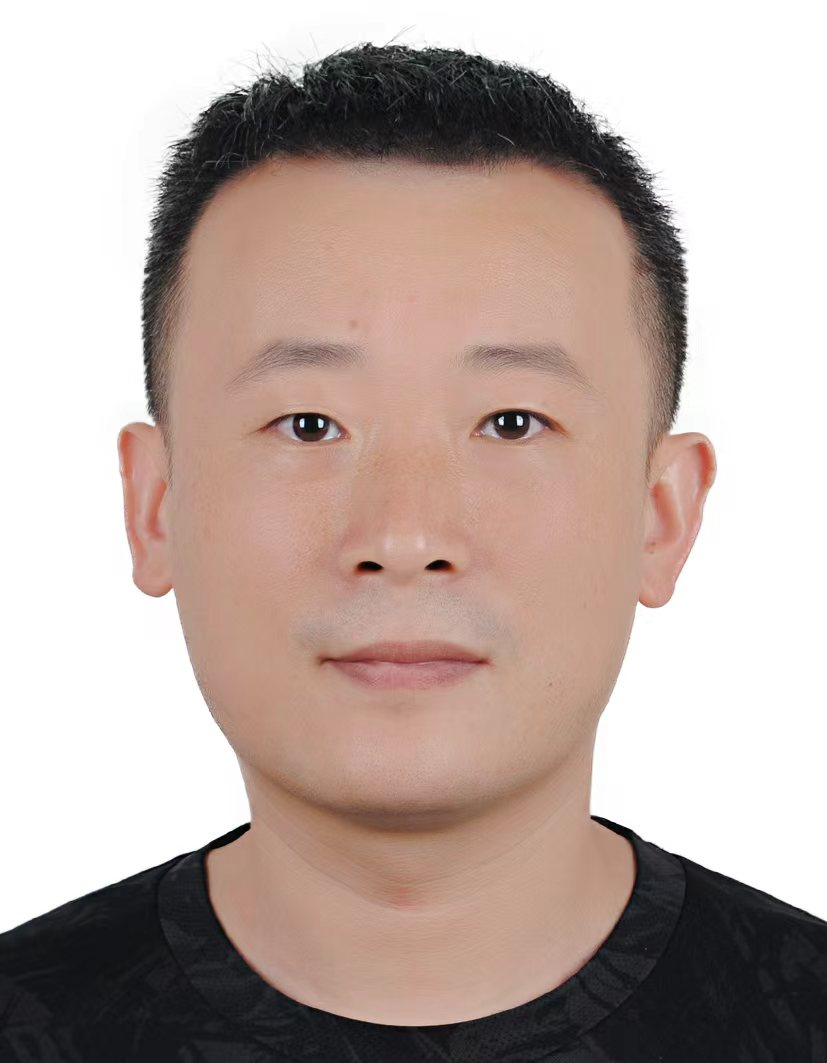}}]{Chaohou Liu}
received the B.S. degree in electrical engineering and automation from Qingdao Agricultural University, Qingdao, China, 
and the M.S. degree in power electronics and electric drive from Shanghai Maritime University, Shanghai, China, in 2008, and 2010, respectively.
He is currently a software leader with SolaX Power Network Technology (Zhejiang) Co., Ltd., Hangzhou, China.

His research interests include micogrid system, energy storage solutions and digital control in power electronics.
\end{IEEEbiography}

\begin{IEEEbiography}[{\includegraphics[width=1in,height=1.25in,clip,keepaspectratio]{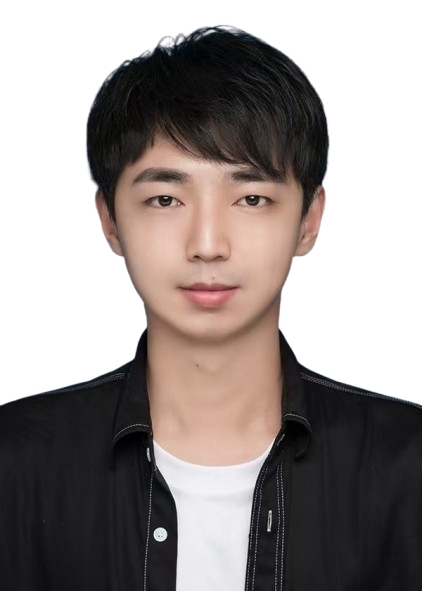}}]{Yang Yang}
received the B.S. degree and the M.S. degree in electrical engineering from Hangzhou Dianzi University, Hangzhou, China, in 2015, and 2018, respectively.
He is currently a product leader with SolaX Power Network Technology (Zhejiang) Co., Ltd., Hangzhou, China.

His research interests include modeling, and design of DC/DC converters and inverters 
for both residential and commercial applications.
\end{IEEEbiography}

\begin{IEEEbiography}[{\includegraphics[width=1in,height=1.25in,clip,keepaspectratio]{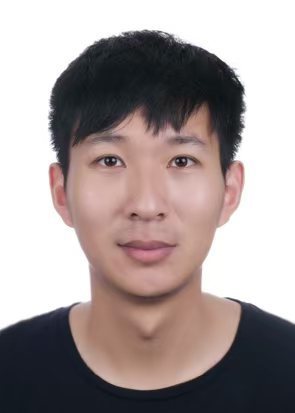}}]{Jiaxi Yin}
received the B.S. degree in Measurement and Control Technology and Instruments 
from Beijing University of Chemical Technology, Beijing, China, in 2013.
He is currently a senior software engineer with SolaX Power Network Technology (Zhejiang) Co., Ltd., Hangzhou, China.

His research interests include embedded system, and control of inverters.
\end{IEEEbiography}

\begin{IEEEbiography}[{\includegraphics[width=1in,height=1.25in,clip,keepaspectratio]{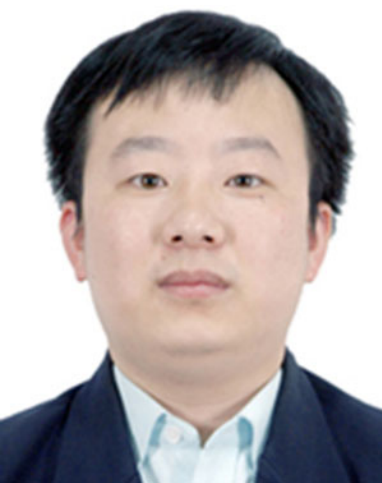}}]{Zeng Liu}
 (Member, IEEE) received the B.S. degree
from Hunan University, Changsha, China, and the
M.S. and Ph.D. degrees from Xi'an Jiaotong University (XJTU), Xi'an, China, in 2006, 2009, and 2013,
respectively, all in electrical engineering.

He then joined XJTU as a Faculty Member in electrical engineering, where he is currently an Associate Professor. 
From 2015 to 2017, he was with the Center for Power Electronics Systems, 
Virginia Polytechnic Institute and State University, Blacksburg, VA, USA,
as a Visiting Scholar. 
His research interests include control of power systems with multiple converters for renewable energy and
energy storage applications, and small-signal stability of power electronics systems.

Dr. Liu was the recipient of two prize paper awards in IEEE TRANSACTIONS
ON POWER ELECTRONICS, and the CPSS Science and Technology Progress Award. 
He serves as an Associate Editor for the IEEE OPEN JOURNAL OF POWER ELECTRONICS and on the Editorial Board for the ENERGIES,
and served as Secretary-General for 2019 IEEE 10th International Symposium
on Power Electronics for Distributed Generation Systems, and 2020 the 4th International Conference on HVDC.

\end{IEEEbiography}

\begin{IEEEbiography}[{\includegraphics[width=1in,height=1.25in,clip,keepaspectratio]{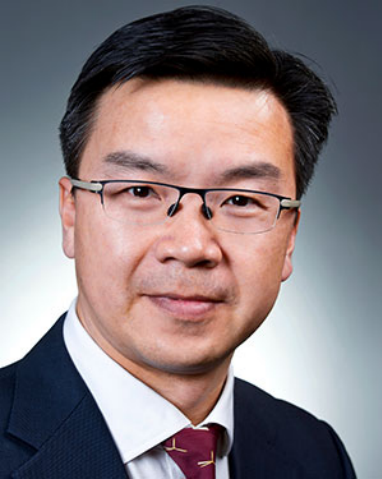}}]{Jinjun Liu}
(M’97–SM’10–Fellow’19) received the B.S. and Ph.D. degrees in electrical engineering from Xi'an Jiaotong University (XJTU), Xi'an, China, in 1992 and 1997, respectively.

He then joined the XJTU Electrical Engineering School as a faculty.
From late 1999 to early 2002, he was with the Center for Power Electronics Systems,
Virginia Polytechnic Institute and State University, Blacksburg, VA, USA, as a Visiting Scholar. 
In late 2002, he was promoted to a Full Professor and then the Head of the Power Electronics and Renewable Energy Center, XJTU, 
which now comprises over 30 faculty members and around 300 graduate students 
and carries one of the leading power electronics programs in China. 
From 2005 to early 2010, he served as an Associate Dean of Electrical Engineering School at XJTU, 
and from 2009 to early 2015, the Dean for Undergraduate Education of XJTU. 
He is currently a XJTU Distinguished Professor of Power Electronics. 
He coauthored 3 books (including one textbook), published over 500 technical papers in peer-reviewed journals and conference proceedings, 
and holds over 90 invention patents (China/US/EU). 
His research interests include modeling, control, and design methods and reliability evaluation and monitoring for power converters and 
electronified power systems, power quality control and utility applications of power electronics, 
and micro-grid techniques for sustainable energy and distributed generation.

Dr. Liu received for many times governmental awards at national level or provincial/ministerial level for scientific research/teaching achievements. 
He also received the 2006 Delta Scholar Award, the 2014 Chang Jiang Scholar Award, 
the 2014 Outstanding Sci-Tech Worker of the Nation Award, 
the 2016 State Council Special Subsidy Award, the IEEE Transactions on Power Electronics 2016 and 2021 Prize Paper Awards, 
the Nomination Award for the Grand Prize of 2020 Bao Steel Outstanding Teacher Award, 
the 2022 Fok Ying Tung Education and Teaching Award, and the 2025 IEEE PELS Harry A. Owen, Jr. Distinguished Service Award. 
He served as the IEEE Power Electronics Society Region 10 Liaison and then China Liaison for 10 years, 
an Associate Editor for the IEEE Transactions on Power Electronics since 2006, and 2015-2021 Vice President of IEEE PELS.
He was on the Board of China Electrotechnical Society 2012-2020 and was elected the Vice President in 2013 
and the Secretary General in 2018 of the CES Power Electronics Society. 
He was 2013-2021 Vice President for International Affairs, China Power Supply Society (CPSS), and since 2016, 
the inaugural Editor-in-Chief of CPSS Transactions on Power Electronics and Applications. 
He was elected the President of CPSS in Nov. 2021. 
Since 2013, he has been serving as the Vice Chair of the Chinese National Steering Committee for College Electric Power Engineering Education Programs.
\end{IEEEbiography}

\end{document}